\documentclass[%
 reprint,
superscriptaddress,
showkeys,
 amsmath,amssymb,
 aps,
prx,
]{revtex4-2}

\usepackage{graphicx}
\usepackage{dcolumn}
\usepackage{bm}
\usepackage{soul}
\usepackage{xcolor}
\usepackage{comment}

\newcommand{\Pe}{\mathrm{Pe}}
\def\NY#1{\textcolor{black}{{#1}}} 
\def\BR#1{\textcolor{black}{{#1}}} 

\begin{document}


\title{Estimation of spatial and time scales of collective behaviors of active matters through learning hydrodynamic equations from particle dynamics}

\author{Bappaditya Roy}
 \email{b.roy@fun.ac.jp }
 \affiliation{
Department of Complex and Intelligent Systems, Future University Hakodate, Kamedanakano-cho 116-2, Hakodate 041-8655 Japan
}
\affiliation{Mathematics for Advanced Materials-OIL (MathAM-OIL), AIST, Katahira 2-1-1, Sendai 980-8577}

\author{Natsuhiko Yoshinaga}
\email{yoshinaga@fun.ac.jp}
\affiliation{
Department of Complex and Intelligent Systems, Future University Hakodate, Kamedanakano-cho 116-2, Hakodate 041-8655 Japan
}
\affiliation{Mathematics for Advanced Materials-OIL (MathAM-OIL), AIST, Katahira 2-1-1, Sendai 980-8577}
\affiliation{WPI-Advanced Institute for Materials Research (WPI-AIMR), Tohoku University, Katahira 2-1-1, Sendai 980-8577, Japan
}%

\date{\today}

\begin{abstract}
We present a data-driven framework for learning hydrodynamic equations from particle-based simulations of active matter. Our method leverages coarse-graining in both space and time to bridge microscopic particle dynamics with macroscopic continuum models. By employing spectral representations and sparse regression, we efficiently estimate partial differential equations (PDEs) that capture collective behaviors such as flocking and phase separation. This approach, validated using hydrodynamic descriptions of the Vicsek model and active Brownian particles \NY{(ABPs)}, demonstrates the potential of data-driven strategies to uncover the universal features of collective dynamics in active matter systems. 
\end{abstract}

\keywords{Active matter; coarse-graining; 
Hydrodynamic equations; Phase transition; Machine Learning}
\maketitle


\section{Introduction}

Collective behavior is a phenomenon in which large groups of individual particles collectively move.
It is ubiquitous for many-body interacting systems.
Even when the interactions are local, collections of the local \BR{interactions} result in large-scale behavior.
When collective behavior occurs, microscopic details of the system become less critical as long as they do not change qualitative features of the system, such as symmetry.
Then, hydrodynamic descriptions give us a universal insight into the system \cite{hohenberg:1977, Marchetti:2013}.
However, the hydrodynamic equations are often hard to derive from microscopic models.
This is because we need a small parameter to realize space and time separation.
Even if such a small parameter is available, the computation to derive the hydrodynamic equations is complex.
Therefore, the hydrodynamic equations are derived from the phenomenological approach in many systems.

For equilibrium systems, we may consider free energy expressed by collective variables.
From the coarse-grained free energy, hydrodynamic equations can be derived from the gradient dynamics minimizing the free energy \cite{onuki:2002}.
This approach cannot be applied to non-equilibrium systems in which a generic description of free energy is not available.
\NY{
For passive systems, we may consider gradient dynamics using the coarse-grained equilibrium free energy under some assumptions or approximations of kinetic coefficients.
For the active system, we may add coarse-grained active terms to the passive dynamics, but they are often introduced based on a sophisticated guess.
}
Then, we have to coarse-grain a microscopic model, such as Langevin dynamics, into hydrodynamic equations expressed by \NY{PDEs}.

\NY{
In the studies of active matters, two major systems, Vicsek model\cite{vicsek:1995,toner:1995, Toner:2012} and \NY{ABPs}\cite{Fily:2012, Cates:2014,Fily:2014a}, were originally proposed as particle models.
}
Their hydrodynamic descriptions are \NY{then} proposed phenomenologically in \cite{toner:1995, Toner:2012} and \cite{Cates:2014}, respectively, based on the symmetry and minimum ways to break equilibrium conditions.
\NY{
After those studies, the connection between them has been studied by deriving the hydrodynamic equations expressed by nonlinear PDEs from the microscopic equations.
These approaches have successfully been applied to
}
the Vicsek model\cite{Bertin:2009,Ihle:2011, Chate:2020} and \NY{ABPs \cite{Speck:2014}}.
\NY{They} originate from the derivation of macroscopic equations of the cytoskeleton from microscopic filament dynamics expressed by the Boltzmann equation\NY{\cite{aranson:2005}} and nonlinear Fokker-Planck equations\NY{\cite{liverpool:2003,kruse:2006}}.
From these studies, we can figure out the origins of phenomenological parameters in the hydrodynamic equations in terms of the parameters in the microscopic models.
A typical assumption made for the derivation is to ignore higher-order correlation in the probability distribution of particle position and orientation.
This is done by taking only two-body collision terms in the Boltzmann equation into account and decomposing the pair distribution into the product of the one-body distribution in the \NY{Fokker-Planck} equation.
In this respect, the derived hydrodynamic equations are not \textit{exact}, particularly at higher particle density, in which many-body correlations come into play\cite{Suzuki:2015}.
Still, the derived hydrodynamic equations capture meso to macroscopic properties of the systems semi-quantitatively. 

The challenge is to develop a systematic method to estimate hydrodynamic descriptions from the microscopic particle dynamics, which is applicable even away from equilibrium systems.
The recent development of machine learning is promising to apply to those problems.
There are two steps for the coarse-graining.
The first step is to identify macroscopic hydrodynamic variables.
The second step is to derive macroscopic governing equations (typically expressed by PDEs) for those variables.
In this study, we focus on the second step.
For the Vicsek \NY{model and ABPs}, density and polarity density are two essential variables because they are conserved or are the variables breaking symmetry.
\NY{
In this study, 
we know those two fields are relevant for the hydrodynamic description.
}
Nevertheless, we will come back to this issue in Sec.~\ref{sec.result.ABP}.

We should stress that our purpose is not to approximate the data of microscopic particle dynamics by macroscopic continuum equations.
In the macroscopic description, details of microscopic dynamics are dropped, and only the \textit{relevant} information, which we expect is universal, is extracted.
Therefore, our approach is entirely different from recently well-studied problems of reduced-order descriptions\NY{\cite{Hijazi:2023}}. 
In their study, the original dynamics are accurately approximated by low-dimensional descriptions to reproduce them.
In this case, the estimation error can be measured by how the solutions of the reduced description deviate from the original dynamics.
On the other hand, in our case, we do not know the ground truth a priori; the hydrodynamic description is convenient for understanding the mechanism of collective behaviors, not reproducing the microscopic details of the dynamics.
The biggest problem is that we do not know how much microscopic information we should remove.
This implies that it is important to clarify which length and time scales we should keep in the hydrodynamic description.
Our study is motivated by this issue.
\NY{
We also point out that recently developed physics-informed neural network (PINN) primarily focuses on solving a given PDE using a neural network instead of using conventional numerical algorithms \cite{Raissi:2019}.
Our purpose is to find a PDE governing coarse-grained scale for a given microscopic particle dynamics, \BR{which} is different from PINN.
}

Estimation of PDEs from data is a recent growing area of machine learning.
Despite the estimation of ordinary differential equations that have often been studied in time-series data analyses, the estimation of PDEs and space-time dynamics looks much harder.
Several machine-learning techniques have been proposed recently, such as sparse identification of nonlinear dynamics (SINDy)\NY{\cite{Brunton2016,Rudy:2017}}, Bayesian modeling for pattern formation\cite{Yoshinaga:2020}.
In fact, these techniques have been successfully applied to pattern formation of block copolymers\cite{Yoshinaga:2020}, the phase-field model for crack propagation\cite{Gao:2022}, and active nematic hydrodynamics estimated from experimental data\cite{Joshi:2022}.


Here, we propose a method to estimate the hydrodynamic equations for two major models in active matter: the Vicsek model and \NY{ABPs}.
We compare the estimated models with the known hydrodynamics equations proposed phenomenologically or derived from the microscopic models.
Our approach is philosophically similar to the studies in \cite{Supekar:2021,Maddu:2022}, in which PDEs are estimated from the data of active matter models.
The advantage of studying these models is that (approximated) hydrodynamic equations derived from the microscopic model are available.
We also focus on the role of coarse-graining in space and time for the estimation results. 

\section{Methods}
\subsection{\BR{Particle} dynamics simulations}
\label{sec.simulations}

This study considers two microscopic models: the Vicsek model and \NY{ABPs}.
In both models, the system is expressed by $N$ particles residing in a two-dimensional square box of size $L$ with periodic boundary conditions. 
Each particle has a position $\mathbf{r}_i$ and a direction $\theta_i$ (or a polarization vector $\mathbf{p}_i = (\cos \theta_i, \sin \theta_i)$). 
\NY{
$N$ particles are initially randomly distributed and oriented on a domain.
}
Each particle exhibits self-propulsion in the direction of $\mathbf{p}_i$ with its speed $v_0$ when isolated.
The Vicsek model has the alignment interaction between particles, and it is known to show global polar order, polar bands, cross-sea, and disordered phases depending on the particle number density and noise\cite{Chate:2020,Kuersten:2020}.
The ABPs focus on the excluded volume interaction between particles, showing motility-induced phase separation (MIPS); namely, the system shows phase separation when the self-propulsion speed (or the Peclet number $\mathrm{Pe}$ defined below) is large\cite{Fily:2012,Redner:2013}.

\subsubsection{Vicsek Model}
We consider a modified version of the Vicsek model\cite{Zhao:2021}, where the dynamics of the particles  ($r_i$, $\theta_i$ for $i=1,2,...,N$) are described as follows:
\begin{align}
\dot{\mathbf{r}}_i &= v_0 \NY{\mathbf{p}_{i}},
\label{emap1}
\\
\dot{\theta}_i &= \kappa \sum_{i \neq j}^{N} F(\theta_j - \theta_i, \mathbf{r}_j - \mathbf{r}_i) + \eta \xi_i(t).
\end{align}
Here, $\kappa$ controls the strength of alignment. $\eta$ accounts for the amplitude of fluctuations or noise.
$\xi_i(t)$ represents Gaussian white noise with zero mean and unit variance acting on the $i$th particle.
The function $F$ describes alignment interactions between particles. For simplicity, we choose $F(\theta, r) = \frac{\sin(\theta)}{\pi R^2} \quad \text{if} \quad |r| < R $ where $R = 1$ is the range of the interaction.
This simple choice for $F$ enforces polar alignment among the particles. While our model is based on this specific form, we expect our results to extend to more general forms of polar alignment. 

We choose the unit of length and time in the simulations of the Vicsek model to be $R$ and $\NY{\tau=}R^2/\kappa$, respectively.
Therefore, the unit length scale is the interaction range between two particles, and the unit of time scale is the time of local alignment.
We set the parameters as $N=10000$, $L=70.6$, $v_0=0.5$, \NY{$\kappa=1$}, and the simulation time step to $0.01\tau$ unless otherwise stated. The number density of the system is $\rho_0=2$. Previous studies have shown that numerical simulations of the Vicsek model exhibit a phase transition between a disordered state and an ordered state as the noise intensity $\eta$ is varied\cite{Chate:2020}. 
The ordered state corresponds to the collective motion with the global polar alignment of the agents, while the disordered state corresponds to the random motion of the agents. Noise \NY{amplitude $\eta$} is a crucial parameter in the Vicsek model as it determines the balance between randomness and the polar order. 

\begin{figure*}[tb]
    \includegraphics[width=1.0\linewidth]{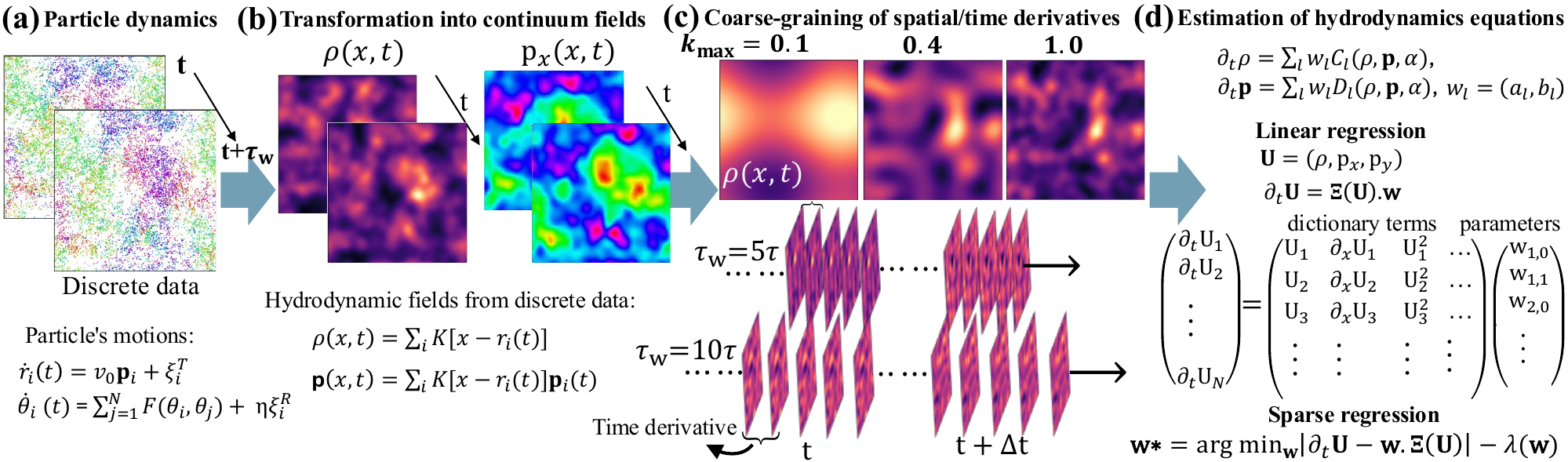}
  \caption{\label{fig.method.schematics} (Color online) Schematic illustration of estimation of hydrodynamic equations from particle dynamics.}
\end{figure*}

To collect rich information on various states, we obtain particle dynamics data under the alternating change of noise strength.
First, we randomly select $10$ values of $\eta$ in $[0.1, \eta_c]$ and $10$ values in $(\eta_c, 0.7]$ to capture the transition between ordered and disordered states.
Here, $\eta_c \approx 0.515$ is the noise strength at the transition.
Then, we vary the noise strength in a piecewise \NY{constant} way. At every $100\tau_w$, we choose $\eta$ alternatively from a value above and below $\eta_c$.
In total, we perform a simulation for $2000\tau_w$ and collect $400$ sampling points in time with $\Delta t=5 \tau_w$.
\NY{
Here, $\Delta t$ is the time window between two sampling points, whereas $\tau_w$ is the time scale to evaluate a time derivative, which eventually sets the time scale of coarse-graining (see Fig.~\ref{fig.method.schematics} and Sec.~\ref{sec.learning.hydrodynamic.equation})
}

\subsubsection{Active Brownian particles}
The dynamics of active particles follow overdamped Langevin equations:
\begin{align}
\NY{
\dot{\mathbf{r}}_i 
}
=& 
\NY{
v_0 \mathbf{p}_i
+
\frac{D \mathbf{F}^{ex}_i}{k_B T} 
}
+ \sqrt{2D} \mathbf{\xi}_i^T,
\label{emap2}
\\
\dot \theta_i =& 
\sqrt{2D_r}\xi_i^R.
\end{align}
The excluded-volume repulsive force \NY{$\mathbf{F}^{ex}$} is derived from the WCA potential $V^{ex}=4\epsilon[(\frac{R}{r})^{12}-(\frac{R}{r})^6]$ if $r<2^{1/6}$. 
Diffusion constants follow the Stokes-Einstein relationship, and $\xi_i^T$ and $\xi_i^R$ are Gaussian white noise.

We make the equations dimensionless using $R$ and $k_BT$ as basic units of length and energy, respectively.
The unit of time scale is chosen as $\tau = R^2/D$.
This is the time scale of translational diffusion by which an isolated particle, on average, moves the distance of particle diameter.
The rotational diffusion constant is $D_r=3D/R^2$, and accordingly, the rotational diffusion time is $\tau_r = \tau/3$.
We perform simulations with $N=40000$ active particles and periodic boundary conditions in a box of size $L=220$. 
The simulation time step is $10^{-5}\tau$. 
The system is characterized by the packing fraction $\NY{\phi}=N\pi(R/2)^2/L^2$ \NY{(or the number density $\rho=N/L^2$)} and the P$\acute{e}$clet number $\mathrm{Pe}=v_0\frac{\tau}{R}$. For the \NY{ABPs}, the Péclet number ($\Pe$) is the key parameter that controls the transition from a non-MIPS state to a MIPS state. By fixing the \NY{volume fraction at $\phi_0\approx0.649$ (or the number density at $\rho_0\approx0.826$)}, we can observe the phase transition by varying the $\Pe$ of the system\cite{Redner:2013}. 
The system remains in a disordered state for sufficiently small values of $\Pe$, while for large enough values of $\Pe$, the system exhibits MIPS.  
Under our choice of $\rho_0$, phase separation occurs at $\Pe \approx \NY{50}$.
In our study, we use data of various values of $\Pe$, perform numerical simulations, and analyze the kinetics during the phase transition. 

To include both information on the phase-separated (MIPS) and non-MIPS regimes and the transitions between them, we performed simulations with various Peclet numbers, covering a range below the transition ($\Pe =0, 20, 40\NY{, 45}$) around the transition point ($\Pe = \NY{50, 55, 60}$), and above the transition ($\Pe = \NY{65, 70,} 80, 100, 120$).
\NY{
We have used the data of $\Pe =0, 20, 40,60,80,100,120$ in Sec.~\ref{sec.result.ABP}, whereas we have used the data of all $\Pe$ to evaluate the transition point more quantitatively in Sec.~\ref{sec.ABP.phase.diagram}.
}
The total time of each simulation is $40.4\tau$, and we collect 100 sampling points in time with $\Delta t = 0.404 \tau$.

\subsection{Learning hydrodynamic equation}
\label{sec.learning.hydrodynamic.equation}
We introduce a learning framework that uses microscopic discrete particle data as input and produces     sparse, interpretable, most expected hydrodynamic equations as output (see Fig.~\ref{fig.method.schematics}\NY{(a-d)}). 
\NY{(a)}The discrete data sets of the particle's trajectories ($r_i (t)$, where $i= 1, 2, ...N$) along with particle orientation ($\theta_i(t)$) at discrete times $(t,t+\Delta t,...,t+T\Delta t)$ are collected \NY{(Sec.\ref{sec.simulations})}. 
\NY{(b)}We then transform the discrete data into continuum fields at each snapshot \NY{(Sec.\ref{sec.hydrodynamic.fields})}.
Subsequently, \NY{(c,d)}we can derive hydrodynamic equations for \BR{the} density field $\rho(t, x)$ and polarization density field $\mathbf{p}(t, x)$. This process involves \NY{(d) constructing over-complete dictionary terms encompassing all conceivable right-hand-side terms, aligning with symmetries and conservation laws (Sec.\ref{sec.dictionary.terms}, see also \cite{Supekar:2021}).
}
We numerically evaluate the values of these terms using the field data.
\NY{(c)}At each sampling point, we compute the spatial and time derivatives of the fields.
To compute the time derivative, we use $5$ snapshots at each sampling point.
\NY{
To realize coarse-graining in time, the time derivatives are computed at different time intervals of $\tau_\mathrm{w}$ (See Sec.~\ref{sec.method.coarsegraining}).
The coarse-graining in space is made by using a filter of wavenumbers when the inverse Fourier transformation of each dictionary term is performed.
}
\NY{(d)}The dictionary terms and time derivative obtained from the data are used for the estimation of the coefficients $\mathbf{w}$.
\NY{
To estimate the most relevant terms, we compute the error during the elimination of each dictionary term and analyze when the optimal number of dictionary terms is attained (Sec.~\ref{sec.regression}).
}
We will explain the details of each step in the \BR{following} sections.

\subsection{Hydrodynamic fields}
\label{sec.hydrodynamic.fields}

After collecting all particle trajectories and orientation data, we first transform particle positions and orientations into continuous fields, using a widely used convolution kernel approach\cite{Hastie:2017}. This is the first stage of the coarse-graining process, aimed at discarding microscopic information below the particle length scale. 
For \NY{$\mathbf{r}_i$} and $\mathbf{p}_i$ representing particle positions and orientations, respectively, the density field $\rho(t, x)$ and polarization density field $\mathbf{p}(t, x)$ can be obtained by

\begin{align}
\label{kernel}
    \rho(t,x) = \sum_{i=1}^{N} K[x-\NY{\mathbf{r}}_i(t)] \\
    \mathbf{p}(t,x) =   \sum_{i=1}^{N} K[x-\BR{\mathbf{r}}_i(t)]\mathbf{p}_i(t)
    ,
\end{align}
where $K(x)=\frac{1}{2\pi \sigma^2} \exp[-|x|^2/2\sigma^2]$ is the Gaussian kernel. 
Here, we choose $\sigma=2$, which determines the minimum spatial resolution.
Coarse-graining of the system beyond the length scale of the particle size will be treated in Sec.~\ref{sec.method.coarsegraining}.

\subsection{\NY{\BR{Dictionary} terms}}
\label{sec.dictionary.terms}

\NY{
To estimate the right-hand side of the hydrodynamic equations, we create} a comprehensive set of dictionary terms that adhere to symmetries and conservation \NY{laws}. 
We use parameterized and nonlinear hydrodynamic equations for density $\rho (\mathbf{r},t)$ and polarity density $\mathbf{p} (\mathbf{r},t)$ with possible candidate dictionary terms in the general form \NY{(see also Fig.~\ref{fig.method.schematics}(d))} 
\begin{align}
   \partial_t  \rho = & \sum_l a_l C_l(\rho,\mathbf{p},\alpha)
   \label{eq.density}
   \\
   \partial_t  \mathbf{p} = &  \sum_l b_l D_l(\rho,\mathbf{p},\alpha) 
   \label{eq.polarity}
   \\
   \dot{\alpha} = & 0
\end{align}
Here, $C_l (\rho, \mathbf{p}, \alpha)$ and $D_l (\rho, \mathbf{p},\alpha)$ are the candidate dictionary terms and functions of the fields and their spatial derivatives (See. Eq.~\eqref{eq.densityb} and Eq.~\eqref{eq.polarityb}).
\NY{
We prepare the dictionary terms $D_l(\rho,\mathbf{p},\alpha)$, up to cubic order in polynomials of $\rho$ and $\mathbf{p}$, including the order of spatial derivatives. 
For $C_l(\rho,\mathbf{p},\alpha)$, we include up to cubic order in polynomials of $\rho$ and $\mathbf{p}$, because the second-order spatial derivative enters due to the conservation law.
To respect the conservation law, we construct the dictionary terms $C_l (\rho, \mathbf{p}, \alpha)$, such that they contain only divergence-form terms, such as $\nabla \cdot \mathbf{p}$ and $\Delta \rho$ (Eq.~\eqref{eq.densityb}).
Similarly, we respect the symmetry so that $C_l (\rho, \mathbf{p}, \alpha)$ and $D_l (\rho, \mathbf{p},\alpha)$ are described in a scalar and vector form, respectively (Eq.~\eqref{eq.polarityb}).
}

We also add a parameter controlling the phase transition into the dictionary terms, namely, $\alpha=\eta$ for the Vicsek model and $\alpha= \mathrm{Pe}$ for the \NY{ABPs}.
With this approach, we may estimate not only the dependence of $C_l$ and $D_l$ on the variables $\rho$ and $\mathbf{p}$ but also on the control parameter $\alpha$\NY{\cite{Brunton2016}}. 
 
\subsection{Spatial and temporal derivatives of coarse-grained fields}
\label{sec.method.coarsegraining}

A fundamental task in estimating hydrodynamic equations is the calculation of spatial and temporal derivatives of coarse-grained fields
\NY{
to estimate the coefficients balancing the left-hand side (time derivative) and the right-hand side (dictionary terms involving spatial derivatives).
We compute the time derivatives of the density field and polarization density field using polynomial finite differentiation at a point, considering different time windows $\tau_w$ for the time derivatives. Understanding how derivatives behave over various time scales 
allows for selecting appropriate time scales for coarse-graining.
}
\NY{
To accommodate various time scales in the coarse-graining process, 
we collect the data at different intervals.
We set $\tau_\mathrm{w} \in [0.1\tau, 50\tau ]$ for the Vicsek model and from $\tau_\mathrm{w} \in [10^{-3}\tau, 0.5\tau]$ for ABPs.
Therefore, time scales $\tau_\mathrm{w}$ for coarse-graining are one of the essential parameters of our methods.
}

\NY{
In the same way, we can also implement spatial coarse-graining on each time frame at fixed time scales for every individual dictionary function. Initially, we perform a Fourier transformation on all dictionary terms with a wave vector $\mathbf{k}=(k_x, k_y)$.
We employ filtering techniques in the dictionary function by considering the values of $k=|\mathbf{k}|$ ranging only in $k \in [0,k_{\mathrm{max}}]$.
Our aim is to collect the data in the wavenumber domain only for these particular ranges of $k$ values and discard the rest. For smaller values of $k_{\mathrm{max}}$, this approach captures the dynamics from long wavelengths, removing most of the microscopic information. In contrast, for larger values of $k_{\mathrm{max}}$, it captures all the information at the microscopic level with short-range wavelengths. 
}

\subsection{Estimation of hydrodynamic equations}
\label{sec.regression}

Besides many methods for the data-driven discovery of dynamical systems, SINDy provides a promising way to identify the dynamical equations from data\NY{\cite{Rudy:2017}}.
To discover PDEs from data, we select only the most informative terms about the dynamics.
We use sparse regression to estimate the hydrodynamic equations consistent with the data of coarse-grained fields discussed in the previous section.

To estimate the PDEs of Eq.~\eqref{eq.density} and Eq.~\eqref{eq.polarity}, we perform the following procedure. 
First, we construct a linear relation of $\partial_t \mathbf{U}=\mathbf{\Xi} (\mathbf{U})\cdot \mathbf{w}$, where the left-hand side is the time derivatives of the fields, $\mathbf{U} = \left(\rho, p_x, p_y \right)$. 
The coefficients $\mathbf{w}=\left( a_l, b_l \right)$ are coefficients in the hydrodynamics equations and are to be estimated.
The row direction of the matrix $\mathbf{\Xi} (\mathbf{U})$ indicates the numerical values of the dictionary terms $C_l(\rho,\mathbf{p},\alpha)$ and $D_l (\rho,\mathbf{p},\alpha)$ obtained from data. 
The column direction of the matrix $\mathbf{\Xi} (\mathbf{U})$ indicates different sampling points in space and time.
The data with different values of control parameter $\alpha$ is stacked in the column direction.
For each variable, $\rho$, $p_x$, or $p_y$, the dimension of $\mathbf{\Xi}(\mathbf{U})$ is $N_d \times M$, where $N_d$ is the total number of sampling points (including different parameter values) and $M$ is the number of dictionary terms.
We aim to infer a parsimonious model, making the vector $\mathbf{w}$ sparse. Importantly, every nonzero value of $\mathbf{w}$ corresponds to a term in the hydrodynamic description. Here, the sparse $\mathbf{w}$ means only a few terms are active. 
Therefore, to estimate the hydrodynamics coefficients $\mathbf{w}$, we use the sparse regression algorithm from SINDy\NY{\cite{Brunton2016,Rudy:2017}}.
\begin{align}
\label{reg}
   \mathbf{w}^* = 
   \underset{\mathbf{w}} {\operatorname{argmin}} 
   \left| \partial_t \mathbf{U} - \Xi(\mathbf{U})\cdot \mathbf{w}  \right|^2
   + \lambda (\mathbf{w}),
\end{align}
where $\lambda (\mathbf{w})$ is a regularization term that controls the level of sparsity. 
There are many ways to realize sparsity through the regularization term.
In \NY{\cite{Brunton2016,Rudy:2017}}, sequential thresholding of estimated parameters $\mathbf{w}$ was used.
In this approach, first, the ridge regression $\lambda (\mathbf{w}) = \lambda_0 \left| \mathbf{w} \right|^2$ is used to estimate $\mathbf{w}$.
Then, all the parameters smaller than a certain value are set to 0.
By iterating the process, a few nonzero parameters are estimated.

In this work, we use a slightly different approach motivated by the method by \NY{\cite{Supekar:2021}}.
This method involves iterating through a series of steps based on the number of dictionary terms.
 At each iteration step, first, we use ridge regression with the regularization parameter $\lambda_0=10^{(-5+w_r)}$, where $w_r$ is the sum of the power of fields present in each dictionary term. 
Then, we eliminate one dictionary term with the smallest coefficient in its magnitude, considering it less important for particle dynamics.
Concurrently, we calculate the error after each dictionary term elimination. 
From the way to change the error $E(n)$ as a function of a nonzero dictionary term $n$, we evaluate the optimal number of dictionary terms $n^*$. 
The error is derived from the sum of squared differences between the predicted values ($\mathbf{\Xi}(\mathbf{U})$) and the actual values ($\partial_t \mathbf{U}$). 
Specifically, the error is defined as 
\begin{align}
E(n)
&=
\frac{1}{N_d} \sqrt{\sum_{i=1}^{N_d} \left( \mathbf{\Xi}(\mathbf{U}_i)\cdot \mathbf{w}_i(n)-\partial_t \mathbf{U}_i \right)^2 }.
\label{eq.error}
\end{align}
Here, $n$ represents the iteration step (i.e., the number of remaining dictionary terms), and $N_d$ is the total number of spatio-temporal sampling data points.
Typical dependence of the error on $n$ can be seen in Fig.~\ref{fig.error.num.terms.VM}(b).
It can be observed that moving from right to left on the plot, the error 
remains almost constant as each dictionary term is eliminated. However, after eliminating a certain dictionary term, the error 
suddenly increases. 
We define $n^*$ as the number of nonzero dictionary terms at which the error starts to deviate from its saturation value.
Quantitative definition of $n^*$ is following:
By fitting $E(n)$ using the initial five points from the right and extrapolating it, we obtain a linear regression line (shown in solid red).
We construct two additional dotted lines above and below the regression line, separated by an interval of $\delta E = 0.005$. 
We define $n^*$ when the error crosses the dotted line.
Note that in \NY{\cite{Supekar:2021}}, $d^2 E(n)/dn^2$ was used to define $n^*$.
This approach does not work in our system because the error decreases as a function of $n$ in multiple steps.
We will discuss this issue in Sec.~\ref{sec.discuss.sparsity} \NY{and Sec.~\ref{sec.ABP.sparsity}}.

\NY{
We should note that in Fig.\ref{fig.error.num.terms.VM}, \ref{fig.error.tau.k}, and \ref{fig.ABP.error.n}, we use the normalized error 
\begin{align}
\tilde{E}(n)
&=
\frac{E(n)}{E(0)}
\end{align}
rather than \eqref{eq.error}.
Therefore, the error decreases from 1 to a smaller value as we increase $n$.
Because our regression method assumes the statistical estimation model as $\partial_t \mathbf{U}=\Xi(\mathbf{U}) \cdot \mathbf{w} + \mathrm{noise}$ for the field variables\cite{Bishop:2006}, a smaller error implies the left-hand side of the hydrodynamic equation is close to the right-hand side.
In the case of the standard supervised learning problem, the error may approach zero as the estimated parameters approaching their ground-truth.
In contrast, in our problem, there is no ground-truth; the hydrodynamic equation does not try to reproduce the detailed dynamics of its microscopic counterparts.
In fact, even after coarse-graining in space and time, there always remains noise in the field data, particularly at the disordered state.
Therefore, the error $E(n)$ (and normalized error $\tilde{E}(n)$ as well) does not go to zero.
}

\section{Vicsek Model}
\label{sec.result.vm}

\subsection{\NY{\BR{Estimation} results}}
\begin{figure}[ht]
    \includegraphics[width=0.49\linewidth]{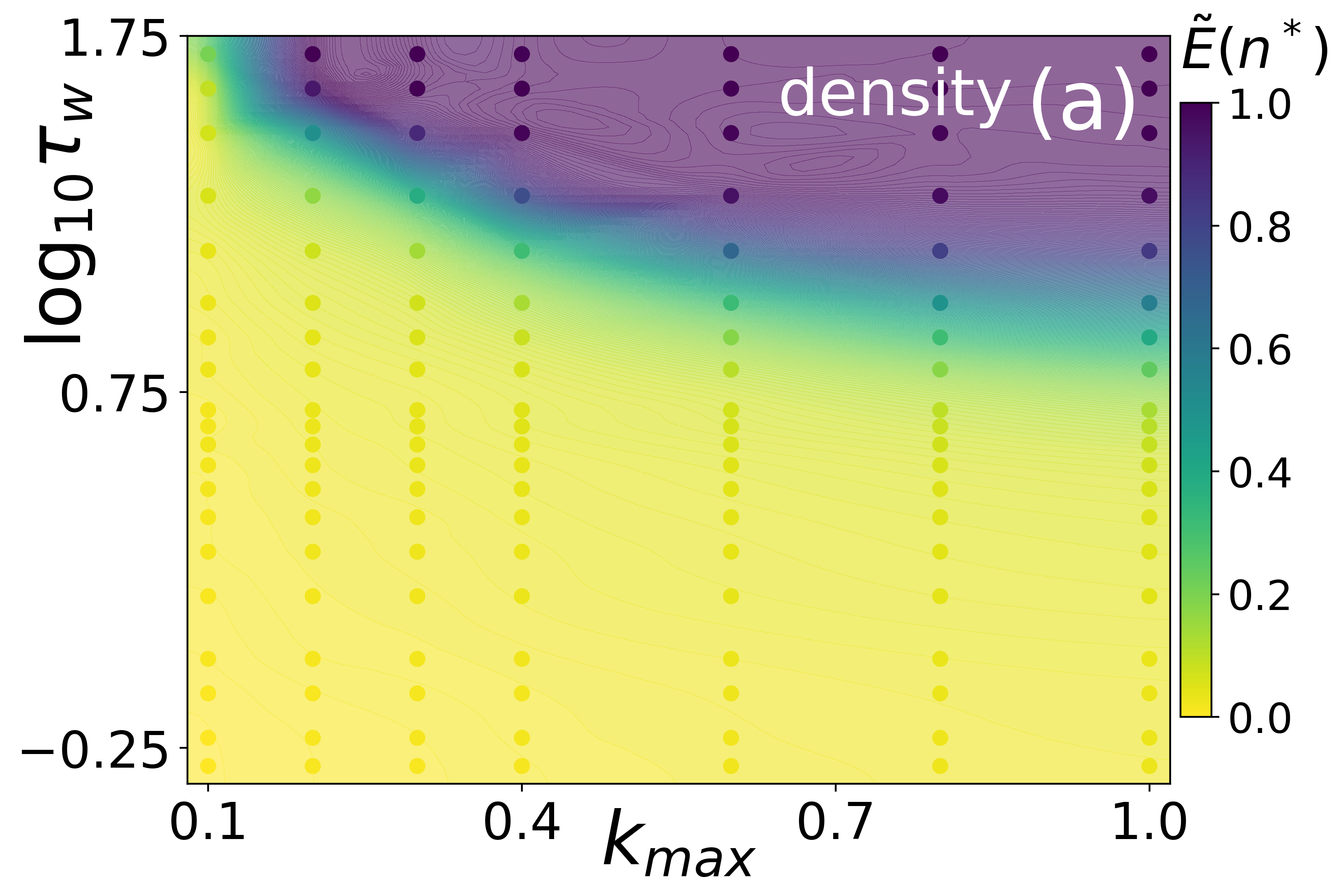}
    \includegraphics[width=0.49\linewidth]{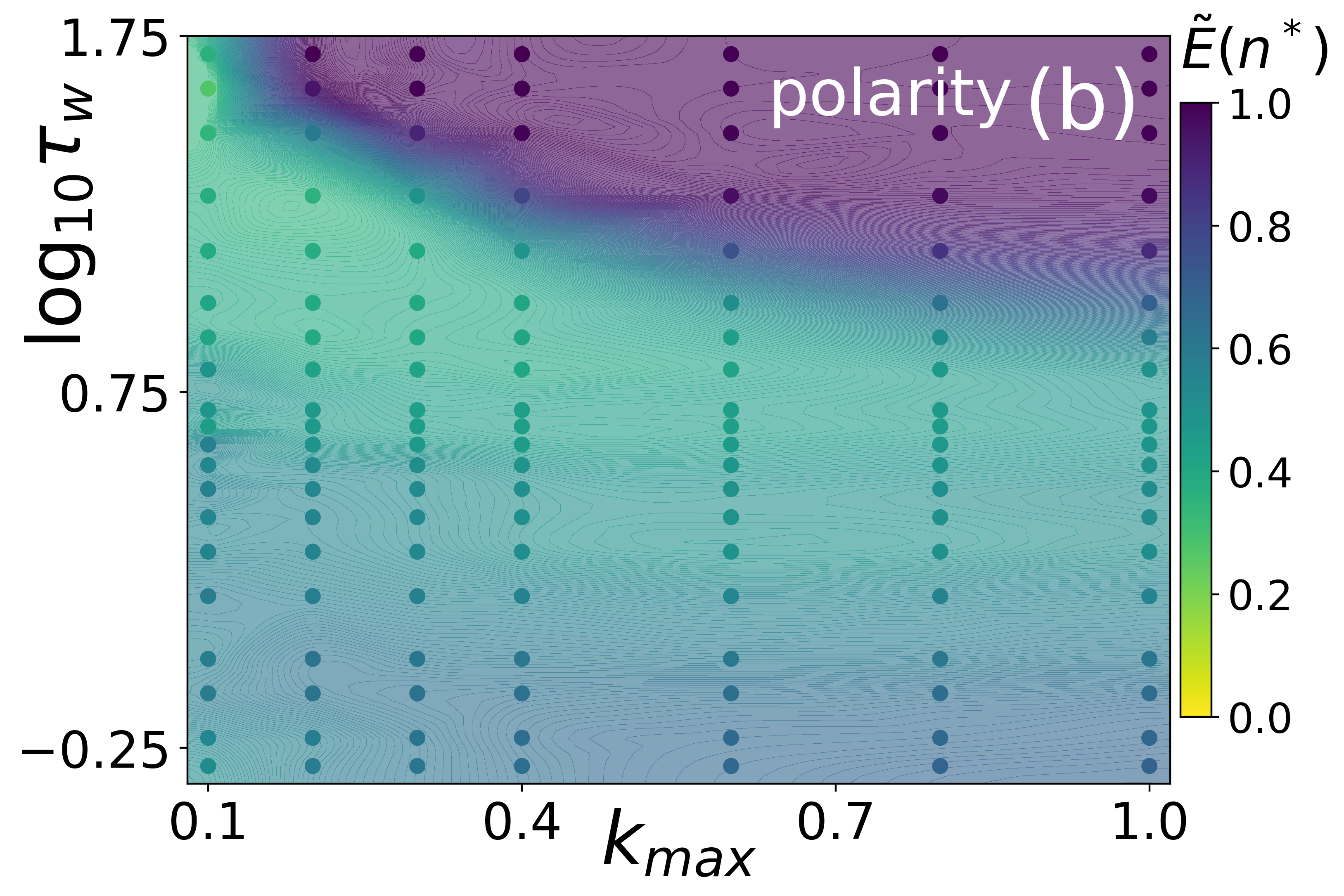}
  \caption{\label{fig.error.tau.k}(Color online) 
  \NY{
  Normalized error $\tilde{E}(n^*)$ at the optimal number of dictionary terms} in ($k_{\mathrm{max}},\tau_\mathrm{w}$) space for density (a) and polarity (b) in the Vicsek model. The color bar represents the change \BR{in} error. 
  }
\end{figure}

Let us start with estimation results for the Vicsek model under specific coarse-graining parameters in space $k_{\mathrm{max}}$ and time $\tau_\mathrm{w}$.
When $k_{\mathrm{max}}=0.1$ and $\tau_\mathrm{w}=40\tau$,
\NY{
the error of the estimation for the polarity density is minimum (see Fig.\ref{fig.error.tau.k} and corresponding texts).
}
We obtain the hydrodynamic equations for density $\rho (\mathbf{r},t)$ and polarity density $\mathbf{p} (\mathbf{r},t)$ \NY{in the form of Eqs.\eqref{eq.density} and \eqref{eq.polarity} as}
\begin{align}
    \partial_t \rho
    =&
    a_1 \nabla \cdot \mathbf{p}
    \label{eq.Vicsek.density}
    \\
    \partial_t \mathbf{p}
    =&
    \NY{\hat{b}_5} \nabla \rho 
    + \NY{\hat{b}_6} (\mathbf{p} \cdot \nabla)\mathbf{p}
        + \NY{\hat{b}_9} \nabla |\mathbf{p}|^2
    + \NY{\hat{b}_{10}} (\nabla \cdot \mathbf{p})\mathbf{p}
    \nonumber \\
    &
    +  
  \NY{\mu (\rho,\eta)} \mathbf{p}
    + 
    \beta (\rho, \eta)|\mathbf{p}|^2 \mathbf{p}
    \label{eq.Vicsek.polarity}
    ,
\end{align}
where $a_1=-0.403$, \BR{$\hat{b}_5=-0.32 + 0.4179\eta$, $\hat{b}_6=0.0051 - 0.3963\eta$, $\hat{b}_9=0.0027+ 0.1769\eta$, and $\hat{b}_{10}=-0.0082 -0.2599\eta$}. 
The coefficients for linear and cubic terms in $\mathbf{p}$ are, respectively, \BR{$\mu(\rho,\eta) = b_3\rho^2 + b_{11} \eta + b_{12} \eta \rho + b_{13} \eta \rho^2$ and $\beta(\rho,\eta) = b_4 + b_{14}\eta$ with $b_3=0.00177$, $b_{11}=-0.00946$, $b_{12}=0.00415$, $b_{13}=-0.00382$, $b_{4}=-0.00177$ and $b_{14}=0.00319$.}
These terms do not contain spatial derivatives, and we call them \textit{bulk terms}.
On the other hand, the terms in the first line of Eq.\eqref{eq.Vicsek.polarity} contain first spatial derivatives.
We call them \textit{advection terms}
\footnote{
The term $\nabla \rho$ is often called a pressure term.
However, we call it an advection term because it arises from the advective part of the Boltzmann equation.
}
.

The \NY{normalized} error in the estimation is small for the equation of density \NY{$\tilde{E}(n^*) \simeq 0.09306$}, whereas for the polarity density, \NY{$\tilde{E}(n^*) \simeq 0.2726$}.
During the estimation process, the error suddenly decreases as in Fig.\ref{fig.error.num.terms.VM}(a) when the term $\nabla \cdot \mathbf{p}$ is included in the hydrodynamic equation for the density field.
Additional terms do not improve the estimation, suggesting that $\nabla \cdot \mathbf{p}$ is the most relevant term in the density dynamics.
Because the density flux is dominated by the flow of self-propelled particles in the Vicsek model, it is reasonable that the flux is proportional to $\mathbf{p}$.
The coefficient of this term is $a_1 \approx 0.4$ in Eq.~\eqref{eq.Vicsek.density}, which is close to the speed of an individual particle $v_0 = 0.5$.
On the other hand, the error in the hydrodynamic equation for the polarity density decreases in multiple steps, as shown in Fig.~\ref{fig.error.num.terms.VM}(b).
The error decreases mostly by the four terms of $\nabla \rho,(\mathbf{p} \cdot \nabla)\mathbf{p}, (\nabla \cdot \mathbf{p})\mathbf{p}, \nabla |\mathbf{p}|^2$.
Still, the error does not reach its minimum value.
The additional \BR{ten} terms decrease the error further.
Beyond \BR{$14$} terms, the error saturates and does not decrease further.
This additional decrease of the error is due to the bulk terms, such as \NY{$ \eta \mathbf{p}, \eta\rho \mathbf{p}, \rho^2 \mathbf{p}, \eta \rho^2 \mathbf{p}, |\mathbf{p}|^2 \mathbf{p}, \eta |\mathbf{p}|^2 \mathbf{p}$, and the advection terms which have different dependence on $\eta$ with the first four terms.}.

\begin{figure}[ht]
    \includegraphics[width=0.49\linewidth]{e-rho-vm-win4th_n.eps}
    \includegraphics[width=0.49\linewidth]{e-pol-vm-win4th_n.eps}
  \caption{\label{fig.error.num.terms.VM}(Color online) 
  \NY{Normalized error $\tilde{E}(n)$} vs $n$ for density field (a) and polarity (b) in the Vicsek model. 
 The optimal numbers of terms are estimated as $n^*=1$ in (a) and $n^*=\NY{14}$ in (b) at $\tau_\mathrm{w}=40\tau$ and $k_{\mathrm{max}}=0.1$.}
\end{figure}

All the estimated terms in Eq.~\eqref{eq.Vicsek.polarity} is coincident with what was studied in the phenomenological hydrodynamic equation\NY{\cite{toner:1995}} and the hydrodynamic equation derived from the Vicsek model\NY{\cite{Bertin:2009}}.
However, the estimated hydrodynamic equation misses several terms.
\NY{
All the diffusive terms in the polarity density, such as $\Delta \mathbf{p}$ and $\nabla (\nabla \cdot \mathbf{p})$
are not estimated.
Note that the anisotropic parts of the diffusive terms
were not obtained by the derivation of the hydrodynamic equation from the Vicsek model\NY{\cite{Bertin:2009}}.
}
\NY{
On the other hand, the diffusive terms may appear when $\tau_w$ is small.
When $\tau_w$ is large, at which the error is low (see Fig.~\ref{fig.error.tau.k}), the data does not contain small-scale inhomogeneity.
The diffusive terms do not appear because they have higher-order spatial derivatives, which scale as $~k^2$.
}

Now, we study how the estimation results depend on coarse-graining length scale $k_{\mathrm{max}}$ and time scales $\tau_\mathrm{w}$.
\NY{
We performed the estimation independently for each $k_{\mathrm{max}}$ and $\tau_\mathrm{w}$, and obtained the optimal number of nonzero dictionary terms $n^*$, their coefficients, and the error $\tilde{E}(n^*)$.
}
For the density field, the error is small, independent of the spatial and time scales, except for large $\tau_\mathrm{w}$ and $k_{\mathrm{max}}$ (Fig.~\ref{fig.error.tau.k}(a)).
The estimated number of terms is always one.
On the other hand, the error for the polarity density is strongly dependent on the spatial and time scales(Fig.~\ref{fig.error.tau.k}(b)).
In fact, the minimum error for the polarity density is attained at a larger length scale $k_{\mathrm{max}} \approx 0.1$ and a longer time scale $\tau_\mathrm{w} \approx 40\tau$.
The results suggest that the hydrodynamic equations are valid only when the spatial scale is larger than $l = 2 \pi /k_{\mathrm{max}} \approx 30R$ and the time scale is longer than $10 \tau$. 
We will compare these length and time scales with the intrinsic scales of the Vicsek model in Sec.~\ref{sec.discussion.scales}.

\begin{figure}[ht]
    \includegraphics[width=0.49\linewidth]{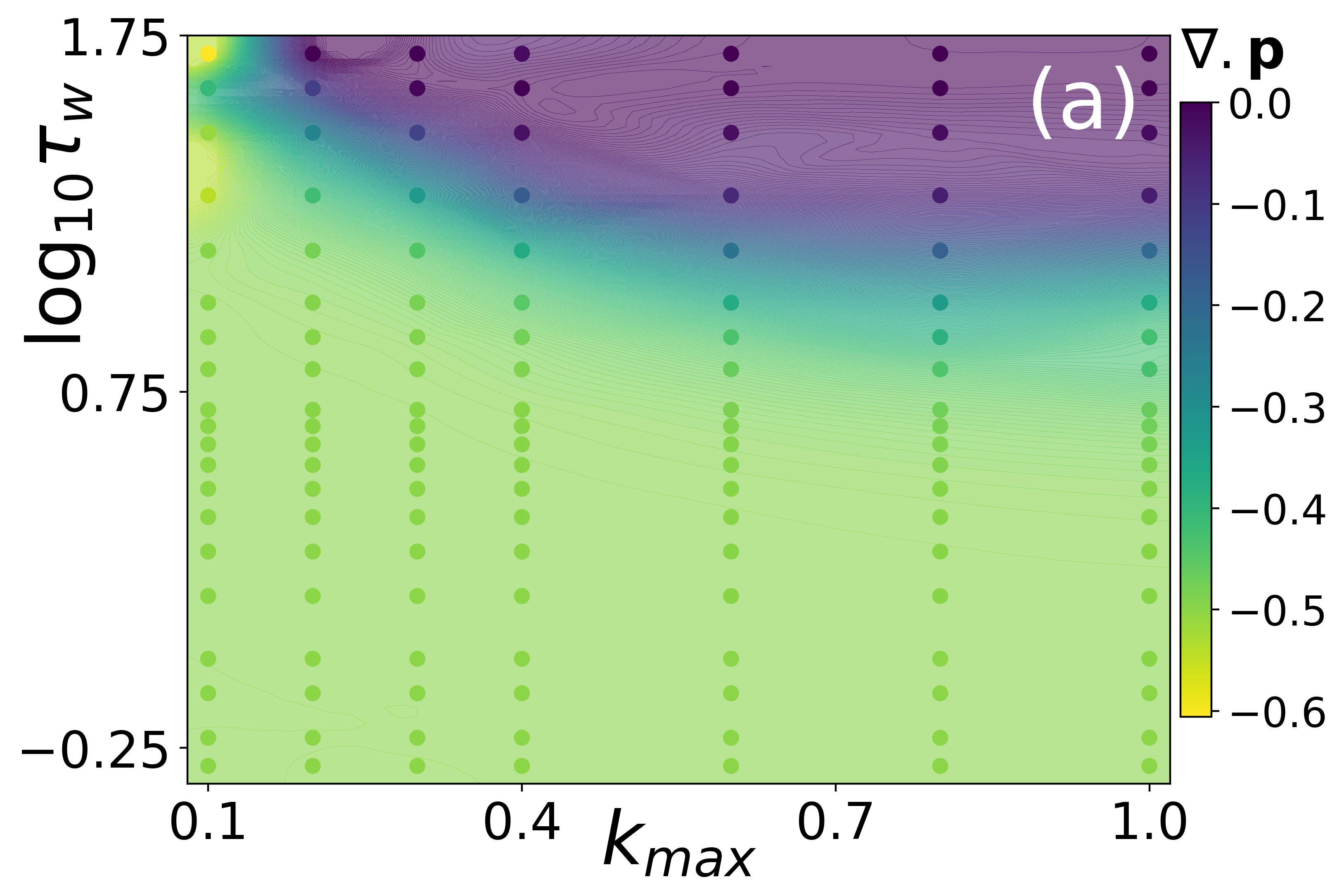}
     \includegraphics[width=0.49\linewidth]{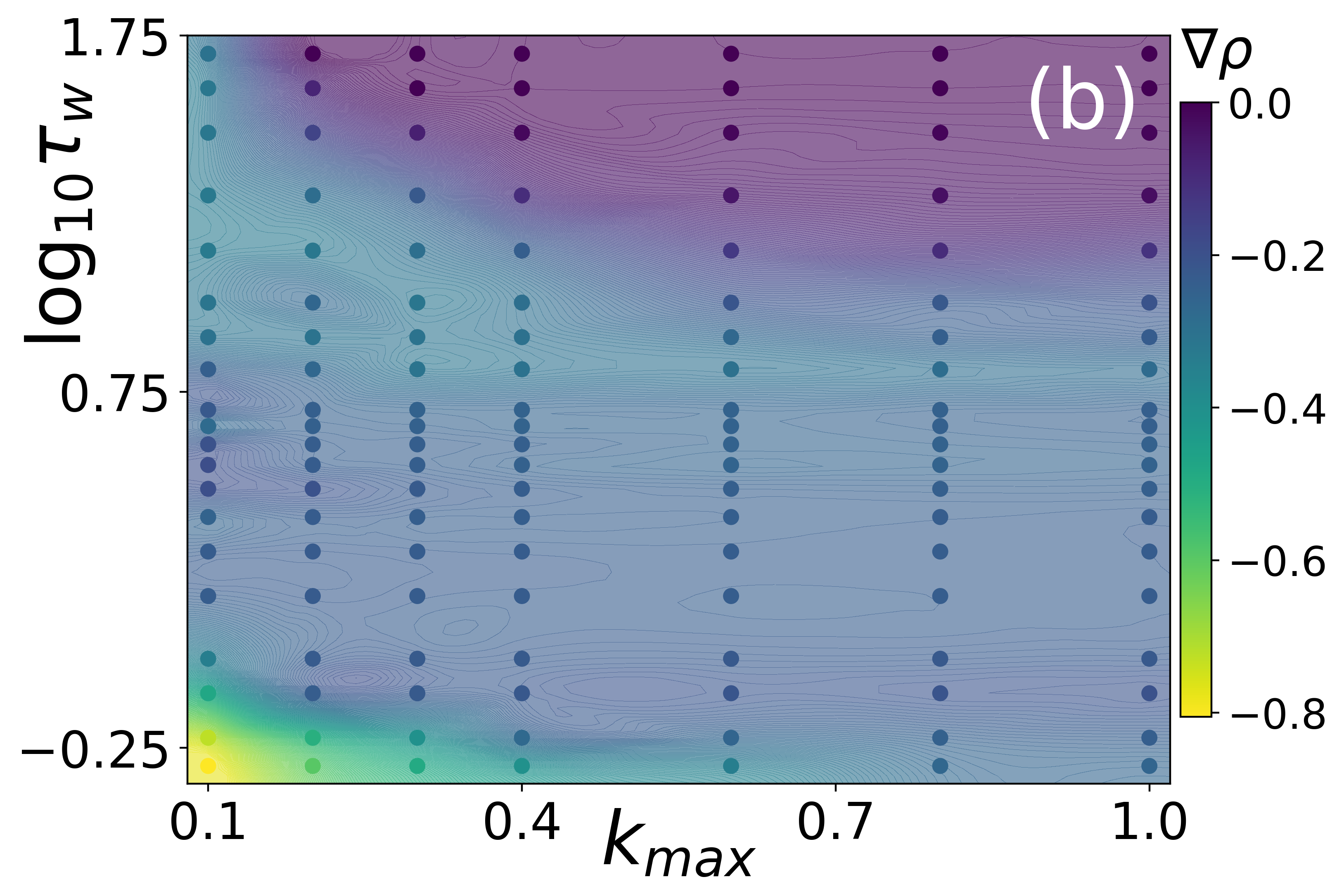}
  \caption{\label{fig.coeff.VM}(Color online) Plot of coefficients of dictionary terms $\nabla \cdot \mathbf{p}$ (a) and $\nabla \rho$ (b) in ($k_{\mathrm{max}},\tau_\mathrm{w}$) space.
  }
\end{figure}

From the results of the error in $(k_{\mathrm{max}}, \tau_\mathrm{w} )$, shown in Fig.~\ref{fig.error.tau.k}, we found that the error of the estimation is dominated by a few terms.
For the hydrodynamic equation of density field, the error decreases as the coefficient of $\nabla \cdot \mathbf{p}$ approaches $-0.5$ (see Fig.~\ref{fig.coeff.VM}(a)).
In fact, the minimum error for the density field is attained at $k_{\mathrm{max}}=0.1$ and $\tau_\mathrm{w}=0.5\tau$.
In this case, the estimated coefficient is $a_1=\BR{-0.499}$ and closer to $-0.5$ than the case of $\tau_\mathrm{w}=40 \tau$ shown above.
Note that $\tau_\mathrm{w}$ at which the error of the density field is minimum is different from $\tau_\mathrm{w}=40 \tau$ at which the error of the polar density is minimum.
Still, the difference in the error of the density field between $\tau_\mathrm{w}=0.5$ and $\tau_\mathrm{w}=40 \tau$ is small.
Therefore, it is reasonable to take $k_{\mathrm{max}}=0.1$ and $\tau_\mathrm{w}=40 \tau$ as an optimal scale of the hydrodynamic equations.

The error of the hydrodynamic equation of polarity density decreases as the coefficient of $\nabla \rho$ approaches $-0.25$.
These results are consistent with those coefficients obtained from the Boltzmann equation: namely, $-v_0$ and $v_0/2$, respectively \NY{\cite{Bertin:2009}}.
Although the smallest error is attained by small $k_{\mathrm{max}}$ and large $\tau_\mathrm{w}$, the error at larger $k_{\mathrm{max}}$ and intermediate $\tau_\mathrm{w}$ is also small.
Both regions have the coefficient of $\nabla \rho$ close to $-0.25$.
The error is smallest at small $k_{\mathrm{max}}$ because the bulk terms can be estimated only in this region.
\NY{
As shown in Fig.~\ref{fig.error.num.terms.VM}(b), the bulk terms make an additional decrease of the error after the appearance of the advection terms (see also Sec.~\ref{sec.discuss.sparsity}).
}

\begin{figure}[ht]
    \includegraphics[width=0.49\linewidth]{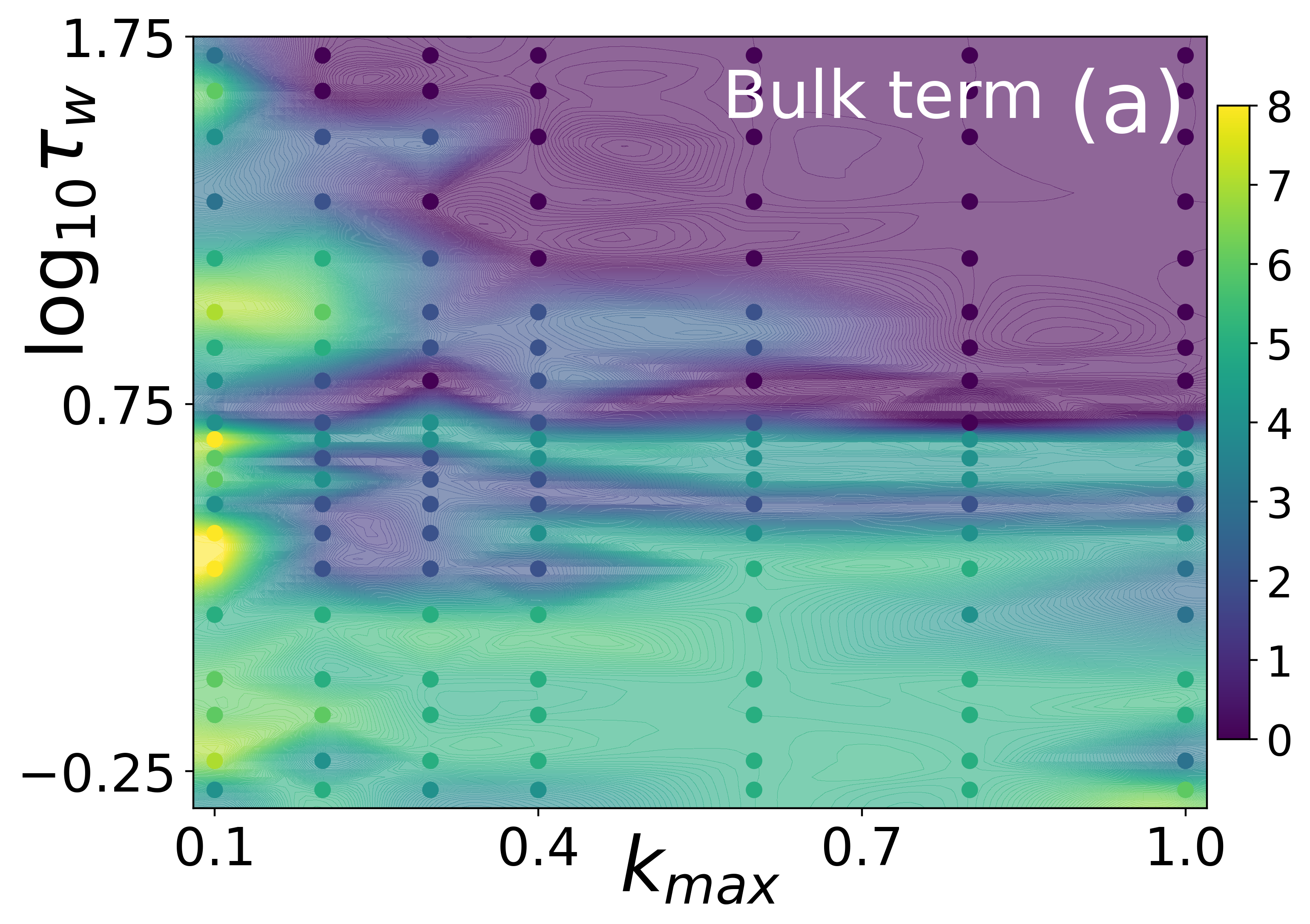}
     \includegraphics[width=0.49\linewidth]{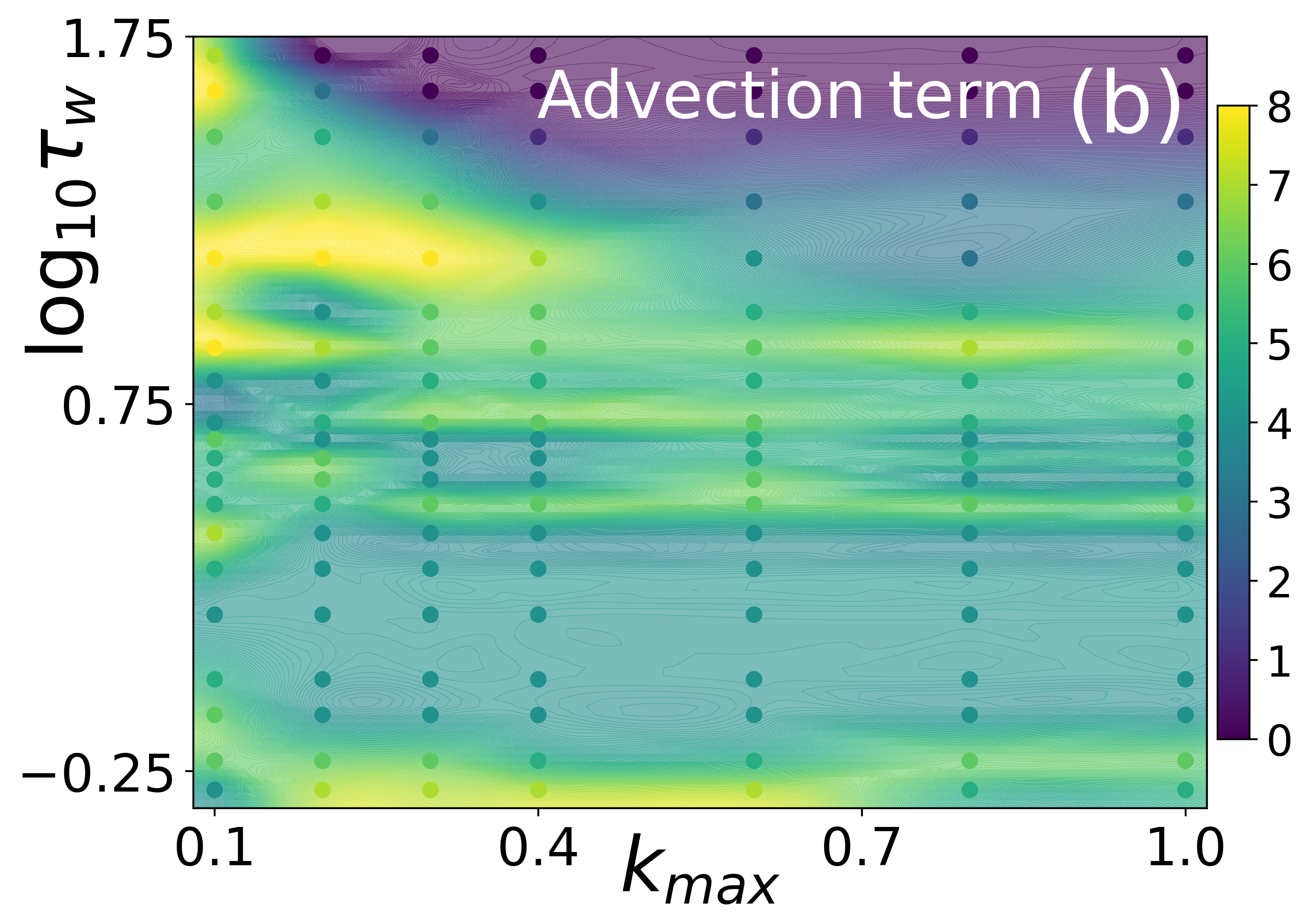}
  \caption{\label{fig.nonzerodictionary.VM}(Color online) Plot of the total number of bulk terms (a) and total number of advection terms (b) in each point of the parameter ($k_{\mathrm{max}},\tau_\mathrm{w}$) space.
  }
\end{figure}

\NY{
For the hydrodynamic equation of polarity density, the estimated number of nonzero dictionary terms and their coefficients are dependent on the spatial and time scales.
In Fig.~\ref{fig.nonzerodictionary.VM}, we show the number of bulk and advection terms at each $k_{\mathrm{max}}$ and $\tau_\mathrm{w}$.
Figure~\ref{fig.nonzerodictionary.VM}(a) shows that the bulk terms, such as $\mathbf{p}$, $\rho \mathbf{p}$, $\rho^2 \mathbf{p}$, $|\mathbf{p}|^2 \mathbf{p}$ and their $\eta$-dependent terms, are estimated only when $k_{\mathrm{max}}$ is small.
On the other hand, 
}
 the four \NY{advection} terms, $\nabla \rho$,$(\mathbf{p} \cdot \nabla)\mathbf{p}$, $(\nabla \cdot \mathbf{p})\mathbf{p}$, and $\nabla |\mathbf{p}|^2 $, are estimated for all time and length scales, as shown in Fig.~\ref{fig.nonzerodictionary.VM}(b).
Because the bulk terms do not contain spatial derivatives, these terms may be comparable with the advection terms only when the wavenumber is small.

\subsection{\NY{\BR{Time} and length scales}}
\label{sec.discussion.scales}

\NY{
In this section, we discuss the length and time scales in the Vicsek \BR{model}.
}
We compare them with the scales on which the estimation results depend.
For the Vicsek model, the time scale unit is chosen by the time scale of alignment $\tau = R^2/\kappa$.
The model has another time scale of collision $\frac{R}{v_0} \sqrt{\frac{\pi}{\rho}}$, which is $\mathcal{O}(\tau)$ with our choice of parameters.
The simulation results show that the time scale of global polarization is $\tau_{p} \approx 200 \tau$.
A small error is attained when $\tau_\mathrm{w} \simeq 10^2 \tau$.
This is comparable with the time scale of global polarization.
In fact, the relaxation time from the disordered state to the ordered state is $40\tau-200\tau$ depending on $\eta$.
The major factor in decreasing the estimation error is advection terms, which play their roles after the polar-ordered state appears.
The length scale of the polar band, which appears near the critical noise strength, is $l \approx 50R$.




\subsection{\NY{\BR{Evaluation} of the estimated results}}
\label{sec.discussion.data.collection}

\begin{figure}[htb]
    \includegraphics[width=0.49\linewidth]{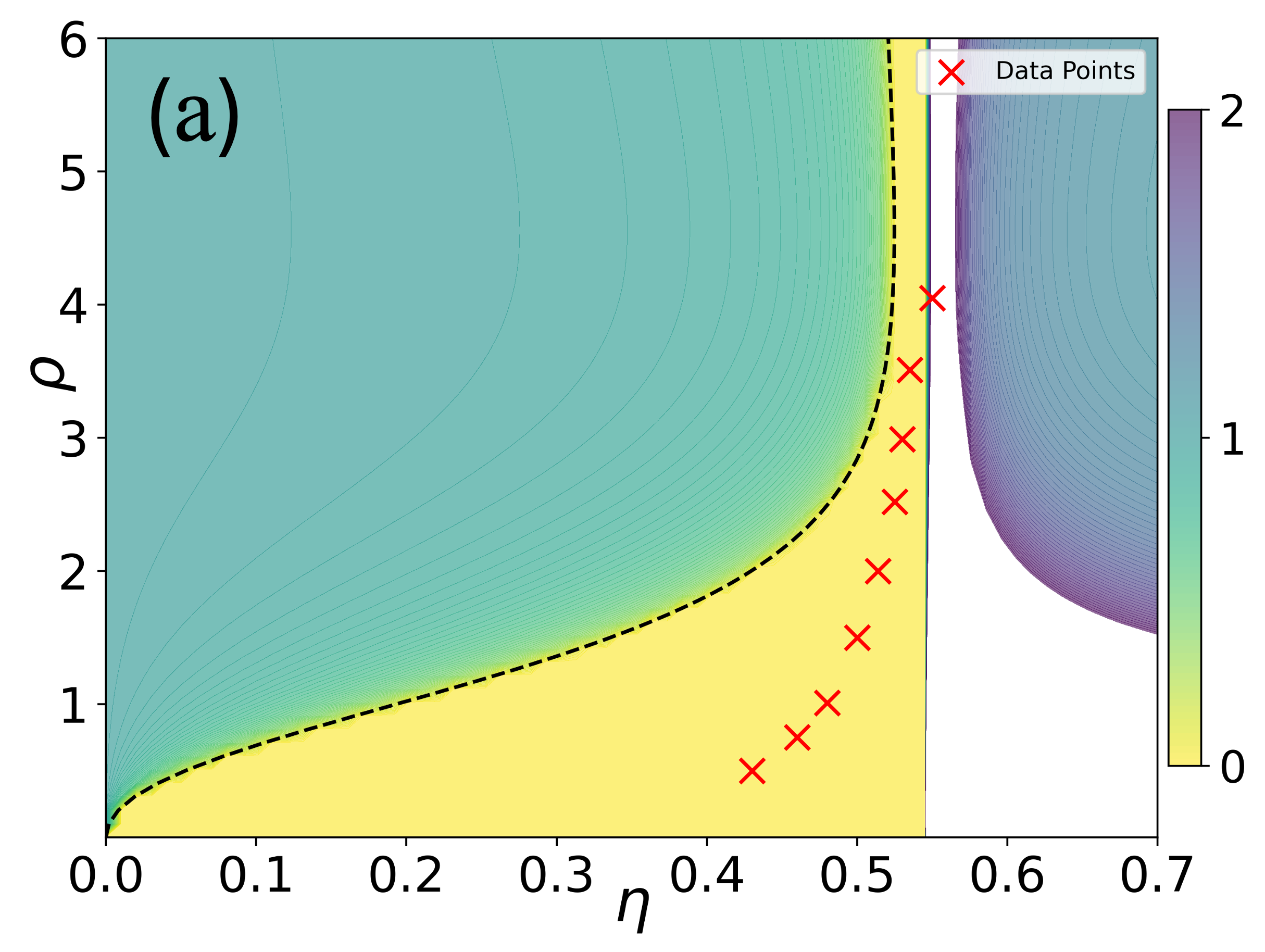}
    \includegraphics[width=0.49\linewidth]{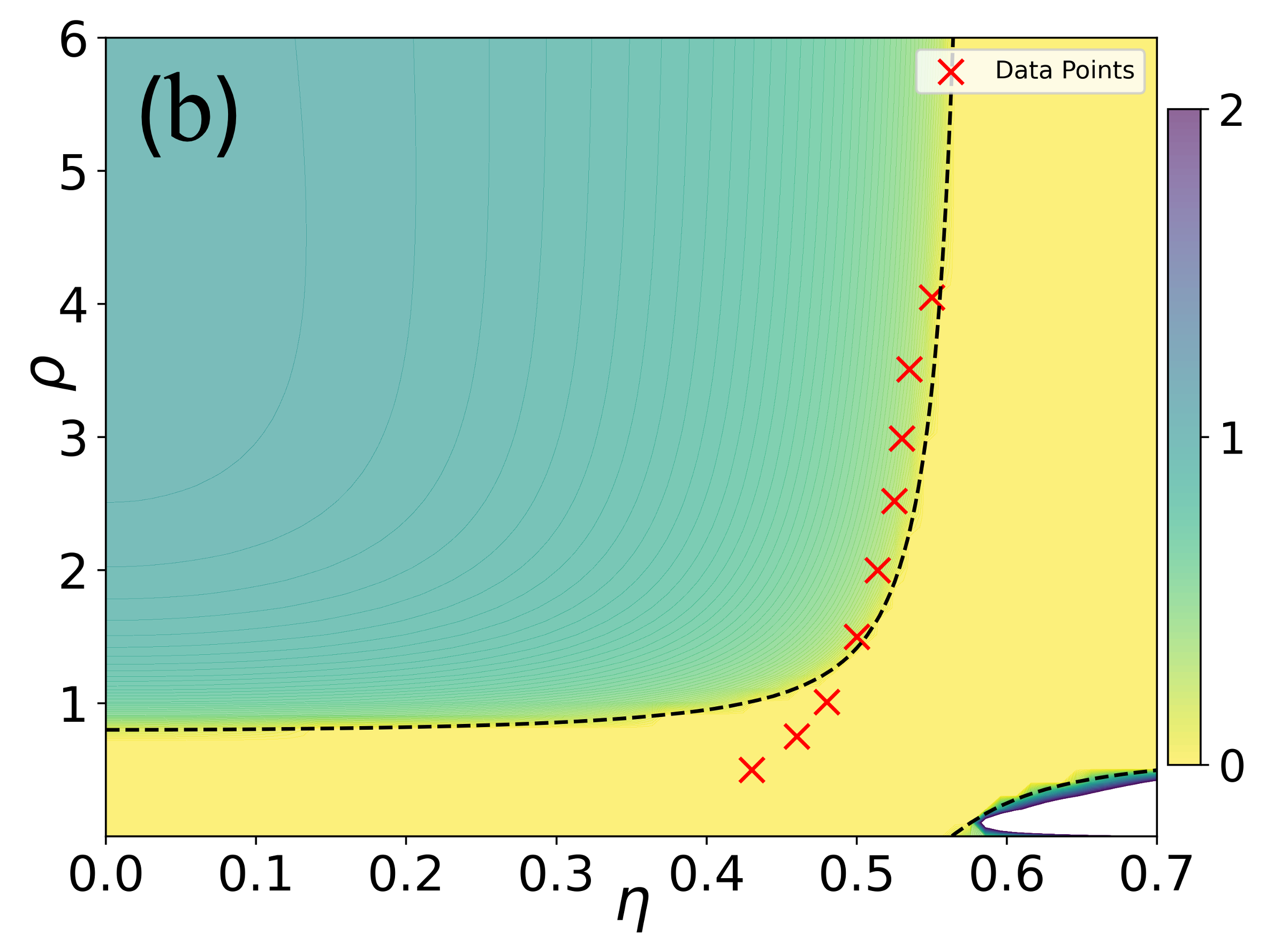}    
  \caption{\label{fig.vicsek.phase}(Color online) 
  Phase diagram showing the transition from the ordered to disordered state in the ($\rho, \eta$) space, evaluated from \NY{$\mu$} and $\beta$ using the estimated coefficients. The black dotted line indicates the phase boundary evaluated by \NY{$\mu(\rho,\eta)=0$}. The color bar on the right represents the magnitude of the polarity $\NY{|\mathbf{p}|/\rho}$.  
  The points shown by the crosses indicate $\eta$ at the phase transition for different values of $\rho_0$ in simulations.
  }
\end{figure}

\NY{
For the validity of the estimation, we check whether the estimated hydrodynamic equation can reproduce the phase diagram of the order-disorder transition of the Vicsek model.
In the Toner-Tu model, the coefficient of the dictionary term $\mathbf{p}$ increases as the noise amplitude decreases and its sign changes from negative to positive.
As a result, the polarity density $\mathbf{p}$ takes a finite value, whose magnitude is determined by $|\mathbf{p}| = \sqrt{-\mu /\beta}$.
As we decrease the amplitude of noise, $\eta$, the sign of $\mu$ changes and, as a result, the phase transition from the disordered state $|\mathbf{p}| =0$ to the ordered state $|\mathbf{p}| \neq 0$ occurs.
}
From the estimated coefficients, we compute the estimated values of $|\mathbf{p}|=\sqrt{-\NY{\mu}/\beta}$ at the steady state.
The estimated \NY{polarity $|\mathbf{p}|/\rho$} as a function of local density $\rho$ and the noise amplitude $\eta$ is shown in Fig.~\ref{fig.vicsek.phase}.
We perform independent simulations of the Vicsek model under the different mean density $\rho_0$ and evaluate the critical value of noise amplitude for each $\rho_0$ (points in Fig.~\ref{fig.vicsek.phase}).
\NY{
We found that the estimated (mean-field) transition line, obtained from $\mu(\rho,\eta)=0$, shows qualitatively the correct dependence on local density and noise.
In this result, the noise-dependence of the coefficient is assumed to be linear in $\eta$.
In \cite{Bertin:2009}, it was suggested that the coefficients of the hydrodynamic equation depend on (powers of) $e^{-\eta^2/2}$ rather than $\eta$.
Therefore, we perform the estimation using $\alpha=e^{-\eta^2/2}$ instead of $\alpha=\eta$ in Eq.~\eqref{eq.polarity}.
The result in Fig.~\ref{fig.vicsek.phase}(b) shows that the phase boundary can be reproduced quantitatively.
}

\subsection{\NY{Generalizability}}
\label{sec.generalizability.VM}

To check whether our estimated hydrodynamic equations are applicable to unseen data, we prepare independent data of microscopic particle simulations and perform a test using the estimated parameters.
\NY{
These data are not used for the estimation of the coefficients.
}
We coarse-grain these \NY{unseen} data and compute the error using the hydrodynamic equations with estimated coefficients for each $n$ \NY{(see Fig.~\ref{fig.error.num.terms.VM}(b)
).
}
We perform the test at least \BR{$12$} times and measure the error at $n^*$ obtained for the training data.
The results are $\NY{\tilde{E}_{\mathrm{test}}}=\BR{0.3317\pm 0.0263}$ for the polarity field of the Vicsek model with $\tau_w=40\tau$ and $k_{\rm max}=0.1$.
These errors are comparable \BR{to} those during training.
The results demonstrate the generalizability of our estimated hydrodynamic equation.

\subsection{\NY{\BR{Sparsity}}}
\label{sec.discuss.sparsity}

When the hydrodynamic equations are estimated using sparse regression, there is always an issue of choosing the optimal number of terms included in the equations.
In this section, we discuss the dependence of the error on the number of nonzero coefficients, which provides information on relevant physical properties in the hydrodynamic description.
Figure~\ref{fig.error.num.terms.discussion} shows \NY{$\tilde{E}(n)$ normalized by its smallest value $\tilde{E}_{\mathrm{min}} = \min_{n} \tilde{E}(n)$ in the Vicsek so that $\tilde{E}(n)-\tilde{E}_{\mathrm{min}}/(1-\tilde{E}_{\mathrm{min}}) \in [0,1]$.
}
We show the results with different $\tau_\mathrm{w}$ but $k_{\mathrm{max}}=0.1$ fixed.
When normalized, all the results show the same trend.
We may see that the error decreases in multiple steps.

For the Vicsek model, the error decreases in two steps at $n_1^*$ and $n_2^*$ (\NY{Fig.~\ref{fig.error.num.terms.discussion}}).
The large part of the error decrease arises from $n_1^*=4$ terms, which are 4 advection terms, $\nabla \rho,(\mathbf{p} \cdot \nabla)\mathbf{p}, (\nabla \cdot \mathbf{p})\mathbf{p}, \nabla |\mathbf{p}|^2$, discussed in Sec.~\ref{sec.result.vm}.
The additional small decrease of the error occurs up to $n_2^*=\NY{14}$.
The additional \NY{10} terms are bulk terms, which are $\mathbf{p}$ and $|\mathbf{p}|^2 \mathbf{p}$ with their coefficients are dependent on $\rho$ and $\eta$\NY{, and advection terms whose dependence on $\eta$ is different from the first $n_1^*$ terms}.
In contrast with the advection terms, these estimated bulk terms differ for different $\tau_\mathrm{w}$\NY{, for example, some $\tau_\mathrm{w}$ includes the term $\rho \mathbf{p}$, whereas others do not.
Still, the tendency of Fig.~\ref{fig.error.num.terms.discussion} is shared by all $\tau_\mathrm{w}$, namely the error drops first by the advection terms and then, a small decrease of error occurs by the bulk terms.
}
This result implies that the hydrodynamic description is dominated by the advection terms, whereas the bulk terms play the role of correction.

\begin{figure}[htb]
    \includegraphics[width=0.7\linewidth]{e-e0-win_n.eps}
\caption{\label{fig.error.num.terms.discussion}(Color online) \NY{
The plot of the error, $\tilde{E}(n)-\tilde{E}_{\mathrm{min}}/(1-\tilde{E}_{\mathrm{min}})$ vs the number of estimated terms $n$ for the Vicsek Model under different time windows $\tau_\mathrm{w}$ with $k_{\rm max}=0.1$.
The error is normalized so that its minimum value is zero for all $\tau_\mathrm{w}$.
  }
  }
\end{figure}

\section{Active Brownian particles}

\subsection{\NY{\BR{Estimation} results}}
\label{sec.result.ABP}

\begin{figure}[ht]
        \includegraphics[width=0.49\linewidth]{e-rho-abp_n.eps}
    \includegraphics[width=0.49\linewidth]{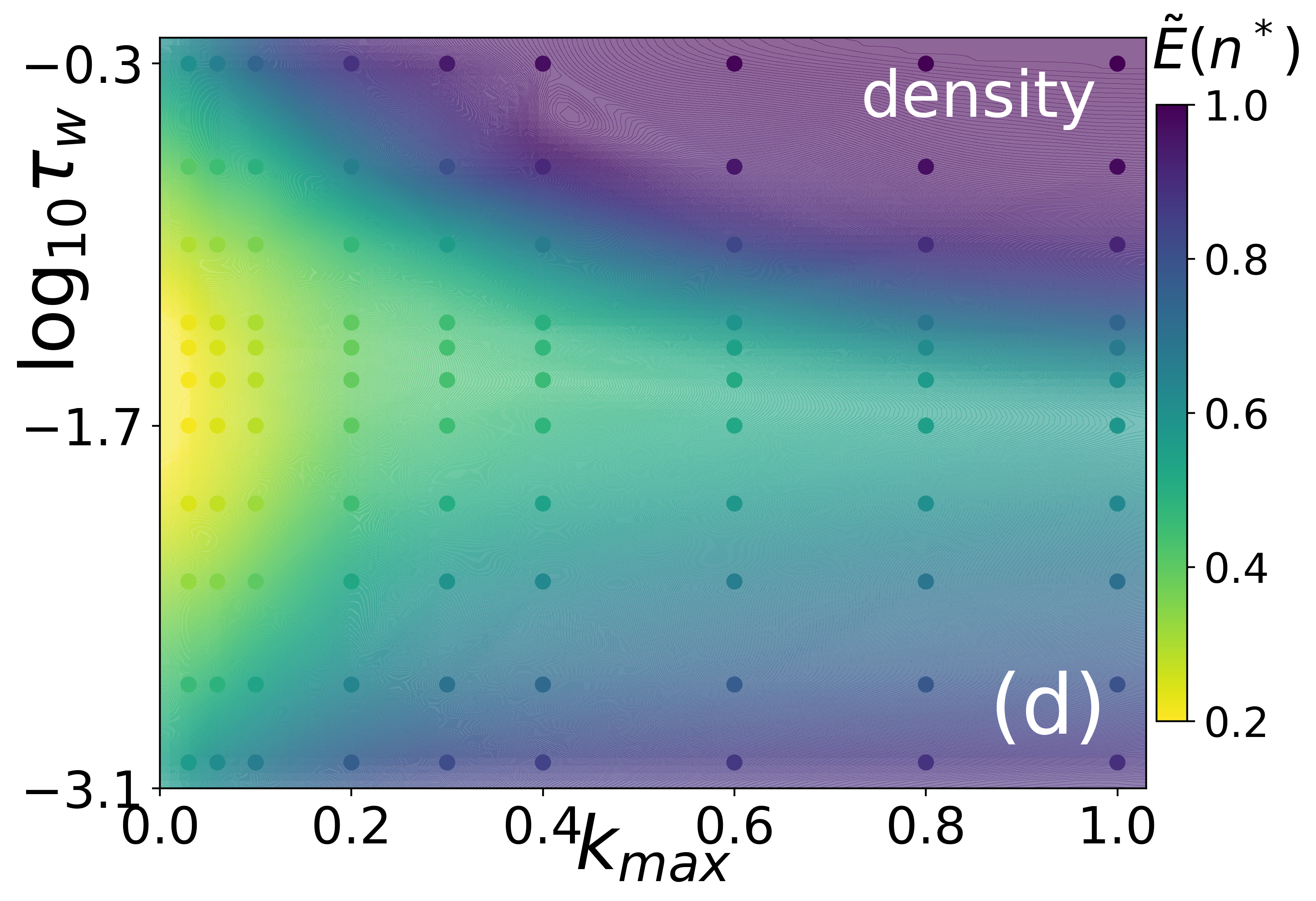}\\
        \includegraphics[width=0.49\linewidth]{e-pol-abp_n.eps}      
    \includegraphics[width=0.49\linewidth]{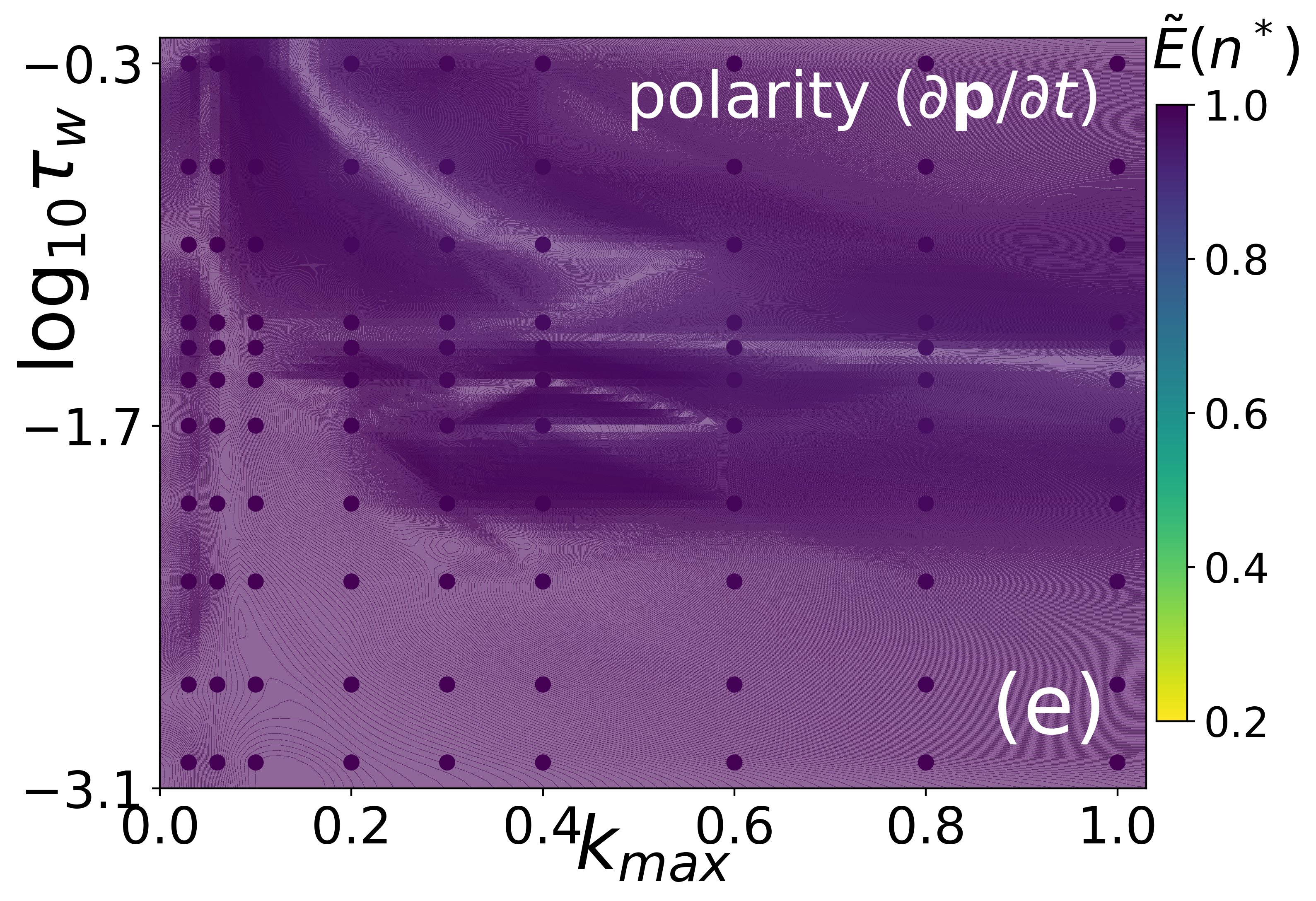}\\
    \includegraphics[width=0.49\linewidth]{e-pg-abp_n.eps}
    \includegraphics[width=0.49\linewidth]{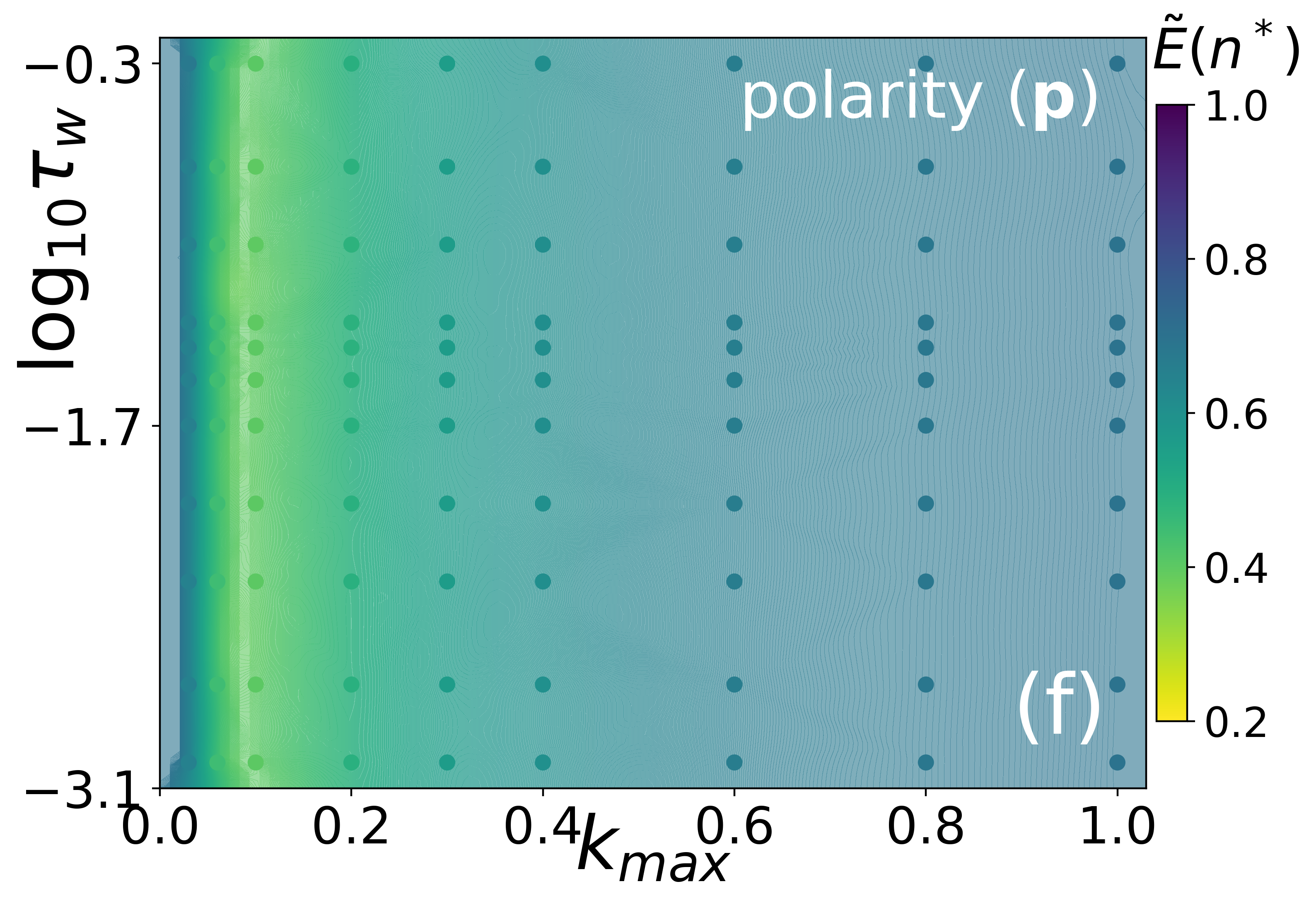}
  \caption{\label{fig.ABP.error.n}(Color online) 
  \NY{(a-c) The normalized error $\tilde{E}(n)$} vs $n$ for \BR{(a) density, (b) polarity ($\partial \mathbf{p}/\partial t$, Eq.~\eqref{eq.polarity}), and (c) polarity ($\mathbf{p}$, Eq.~\eqref{eq.polarity2})} in the \NY{ABPs} at $k_{\mathrm{max}}=0.1$ and $\tau_\mathrm{w}=0.02\tau$. 
  \NY{(d-f) The normalized error $\tilde{E}(n^*)$ at the optimal number of dictionary terms} for density ($\partial \rho/\partial t$) (d), \BR{polarity ($\partial \mathbf{p}/\partial t$, Eq.~\eqref{eq.polarity}) (e), and polarity ($\mathbf{p}$, Eq.~\eqref{eq.polarity2}) (f)} in the $(k_{\mathrm{max}},\tau_\mathrm{w})$ space, respectively.
  }
\end{figure}

In the case of ABPs, our estimation result is\NY{, in the form of Eqs.~\eqref{eq.density} and \eqref{eq.polarity2}, which we will explain later, as}
\begin{align}
    \partial_t \rho
    =&
     \nabla \cdot  \left[
     f_1(\rho, \Pe) \mathbf{p}
     \right]
    + \Delta f_2 (\rho, \Pe)
    + \Delta^2 f_3 (\rho, \Pe)
    + \mathcal{N}[\rho,\mathbf{p}]
    \label{eq.ABP.density}
    \\
    \mathbf{p}
    =&
    \nabla g_1 (\rho, \Pe)
    + \nabla \Delta g_2 (\rho, \Pe)
    \label{eq.ABP.polarity}
    ,
\end{align}
when $k_{\mathrm{max}}=0.1$ and $\tau_\mathrm{w}=0.02\tau$. 
\NY{
For the density field, the estimation error decreases in three steps as we increase the number of terms, as shown in Fig.~\ref{fig.ABP.error.n}(a).
First, the error decreases significantly by $5$ terms, which are $\Delta \rho, \Delta \rho^2, \Pe \nabla\cdot\mathbf{p}, \Pe \Delta \rho^2, \Pe \nabla\cdot(\rho\mathbf{p})$.
Among those terms, $ \Pe \nabla\cdot\mathbf{p}$ and $\Pe \Delta \rho^2$ make the biggest contribution to the decrease of error.
After that, the error stays at a more or less constant value, and then by an additional $7$ term, the error decreases again.
At the constant error around $\NY{\tilde{E}(n)} \approx 0.4$, the advection terms without $\Pe$, such as $\nabla \cdot \mathbf{p}$ and $\nabla \cdot (\rho \mathbf{p})$, and surface terms $\Delta^2 \rho $, $\Delta^2 \rho^2$, $\Pe \Delta^2 \rho$, and $\Pe \Delta^2 \rho^2$ appear.
Then, the error further decreases by $\Pe \Delta \rho^3$.
Finally, the error slightly decreases by the additional $7$ terms, including the non-gradient term $\mathcal{N}[\rho,\mathbf{p}] = 150.371\Delta |\mathbf{p}|^2$, saturates at $\NY{\tilde{E}(n^*)} \simeq 0.2832$, and does not show a further decrease by adding other terms.
}

\NY{On the other hand,} Eq.~\eqref{eq.ABP.polarity} requires an explanation.
First, we estimate the coefficients of the hydrodynamic equation in the form of Eq.~\eqref{eq.polarity}.
The error of this estimation is shown in Fig.~\ref{fig.ABP.error.n}\NY{(b)}.
Even if we include all the dictionary terms, the error remains high, $\NY{\tilde{E}(n)} \simeq 1$.
This result suggests that the polarity density $\mathbf{p}$ is not a hydrodynamic variable. 
The polarity density $\mathbf{p}$ is slaved by the density field $\rho$.
Then, instead of Eq.~\eqref{eq.polarity}, the polarity density should be expressed by the function of the density field.
Therefore, we consider the following estimation:
\begin{align}
    \mathbf{p}
    =& 
    \sum_l \tilde{b}_l \tilde{D}_l(\rho,\alpha) 
     \label{eq.polarity2}
\end{align}
with the coefficients $\tilde{b}_l$ for each dictionary term (see Eq.\eqref{eq.polarity2b}).

The error is shown in Fig.~\ref{fig.ABP.error.n}(c).
In this case, the error decreases in two steps.
First, the error decreases significantly by the term $\Pe \nabla \rho^2$, and then it decreases slightly by the additional $5$ terms.
Finally, the error saturates at $\NY{\tilde{E}(n)} \simeq 0.4$.
The estimated terms are shown in Eq.~\eqref{eq.ABP.polarity}.
The estimated terms, $f_1,f_2,f_3,g_1,g_2$, in Eq.~\eqref{eq.ABP.density} and Eq.~\eqref{eq.ABP.polarity} are polynomials of $\rho$ and linear functions of $\Pe$.
The detailed expressions of the coefficients are shown in Appendix~\ref{appendix.ABP.coeff}.
\NY{
The estimation of the density equation works similarly to the Vicsek model, suggesting that the density is a hydrodynamic variable in this model.
}

\begin{figure}[ht]
    \includegraphics[width=0.7\linewidth]{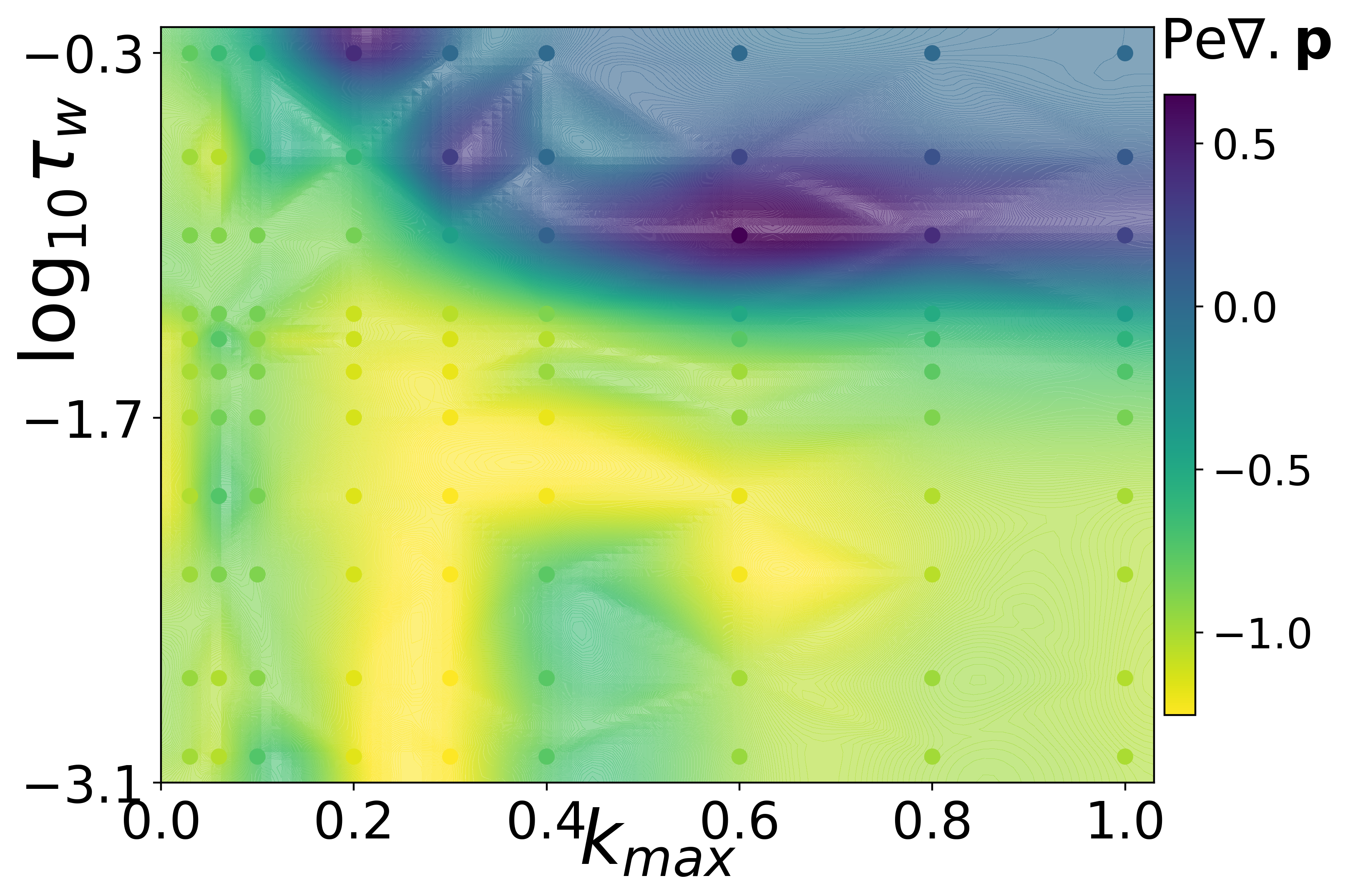}
  \caption{\label{fig.coeff.ABP}(Color online) 
  Plot of coefficients of dictionary terms $\Pe\nabla\cdot\mathbf{p}$ in ($k_{\mathrm {max}},\tau_\mathrm{w}$) space.
  }
\end{figure}

The essential term driving MIPS is $\Pe \nabla \cdot \mathbf{p}$.
Together with the estimation result of $\mathbf{p} = \NY{\tilde{b}_{2}} \nabla \rho^2$ \NY{in Eq.~\eqref{eq.ABP.polarity} (or Eq.\eqref{eq.polarity2})}, the term $a_{14} \Pe \nabla \cdot \mathbf{p}$ \NY{in Eq.~\eqref{eq.ABP.density} (or Eq.\eqref{eq.density})} results in negative diffusion and destabilize the homogeneous state when $a_{14} <0$ and $\NY{\tilde{b}_2}>0$.
The negative diffusion is balanced with the positive diffusion term $\Pe \Delta \rho^2$.
This balance may be the reason for the main decrease in the error by $\Pe \nabla \cdot \mathbf{p}$ and $\Pe \Delta \rho^2$ for the density field.
The coefficient $a_{14} \Pe$ corresponds to the negative sign of the velocity of non-interacting active particles.
In fact, our estimation shows $a_{14} \approx -0.887$, which is close to $-1$ (see Fig.~\ref{fig.coeff.ABP}).

We perform the estimation for different time $\tau_\mathrm{w}$ and length $k_{\mathrm{max}}$ scales.
\NY{
For the dynamic of the density field, the minimum error is attained when $\tau_\mathrm{w} = 0.02 \tau$ and $k_{\mathrm{max}}=0.03$ (Fig.~\ref{fig.ABP.error.n}(d)).
Interestingly, the minimum error appears at the intermediate time scale.
When the time scale is too long, the error starts to increase again.
}
Figure~\ref{fig.ABP.error.n}\NY{(e)} shows that the error of the estimation based on Eq.~\eqref{eq.polarity} is always large, whatever the time and length scales.
This result also supports the above-mentioned argument; the polarity density is not the hydrodynamic variable in this model.
Instead, Fig.~\ref{fig.ABP.error.n}\NY{(f)} shows smaller error at $k_{\mathrm{max}} \approx 0.1$.
\NY{
We should note that for the density field, even though the smallest error is attained at $k_{\mathrm{max}} = 0.03$, the error is also small at $k_{\mathrm{max}} = 0.1$.
On the other hand, the error of the polarity density is high at $k_{\mathrm{max}} = 0.03$.
This is because, when $k_{\mathrm{max}}$ is too small, spatially inhomogeneous structures almost disappear after filtering.
Because the magnitude of the polarity density $\mathbf{p}$ is accumulated at the position where the density field is inhomogeneous, the relation between $\mathbf{p}$ and $\rho$ is not well captured for too small $k_{\mathrm{max}}$.
This is the reason why we choose $k_{\mathrm{max}} = 0.1$ as a representative example of the estimation results at the beginning of this section.
}

The dominant contribution to the small error for the density field of ABPs is $\Pe \nabla \cdot \mathbf{p}$.
When this term approaches $-1$, the error becomes smaller, as seen in Fig.~\ref{fig.coeff.ABP}.
At larger $k_{\mathrm{max}}$ and larger $\tau_\mathrm{w}$, $\Pe \nabla \cdot \mathbf{p}$ deviates from $-1$.
When $\tau_\mathrm{w}$ is smaller than $10^{-2}\tau$, the error increases.
This is because the phase separation occurs at a longer time scale than local cluster formation, which is $\approx \tau /\Pe$.
We compare the time scale appearing in the estimation results and intrinsic time scales of ABPs in Sec.~\NY{\ref{sec.abp.scales}}.

\begin{figure}[ht]
    \includegraphics[width=0.49\linewidth]{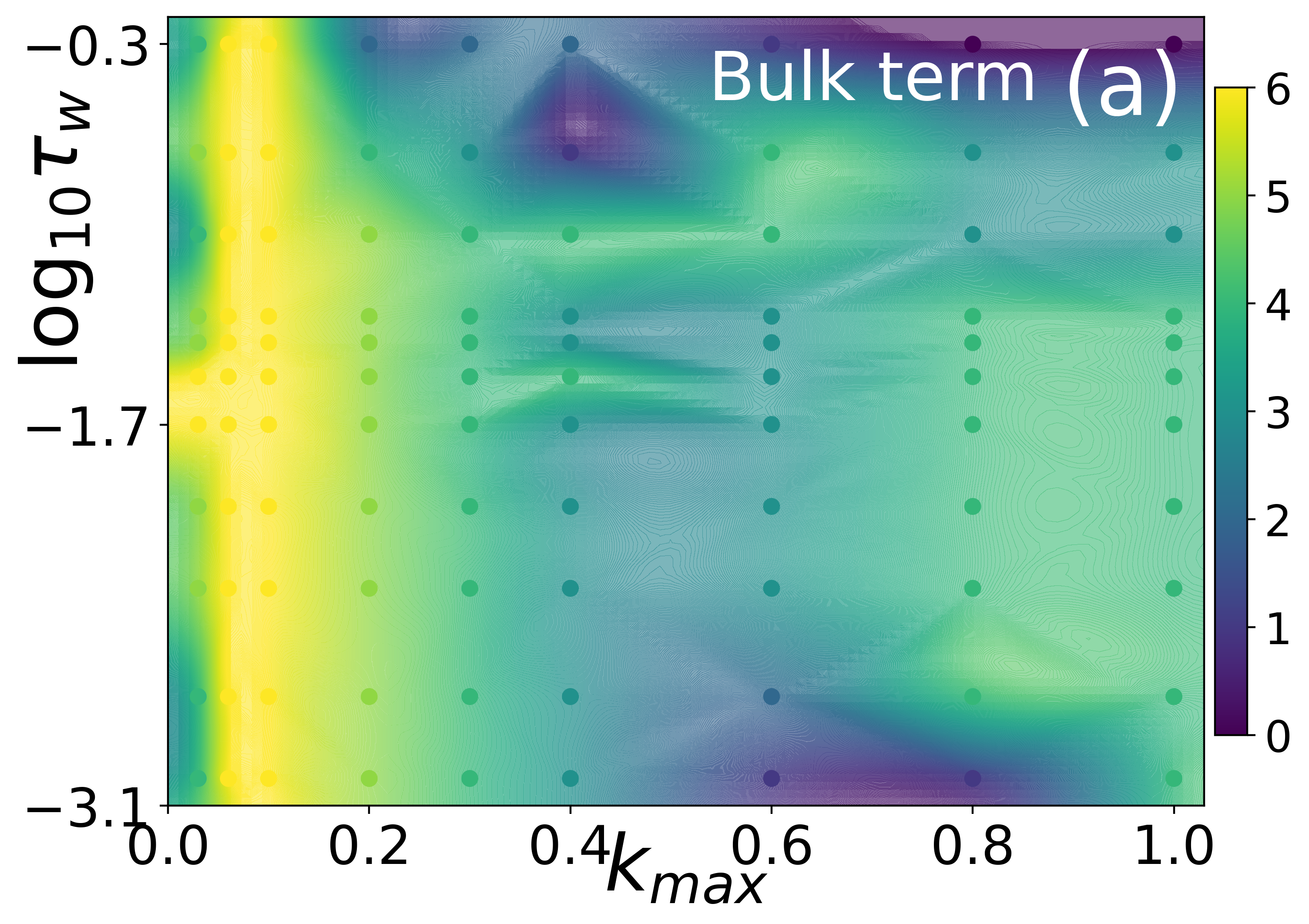}
     \includegraphics[width=0.49\linewidth]{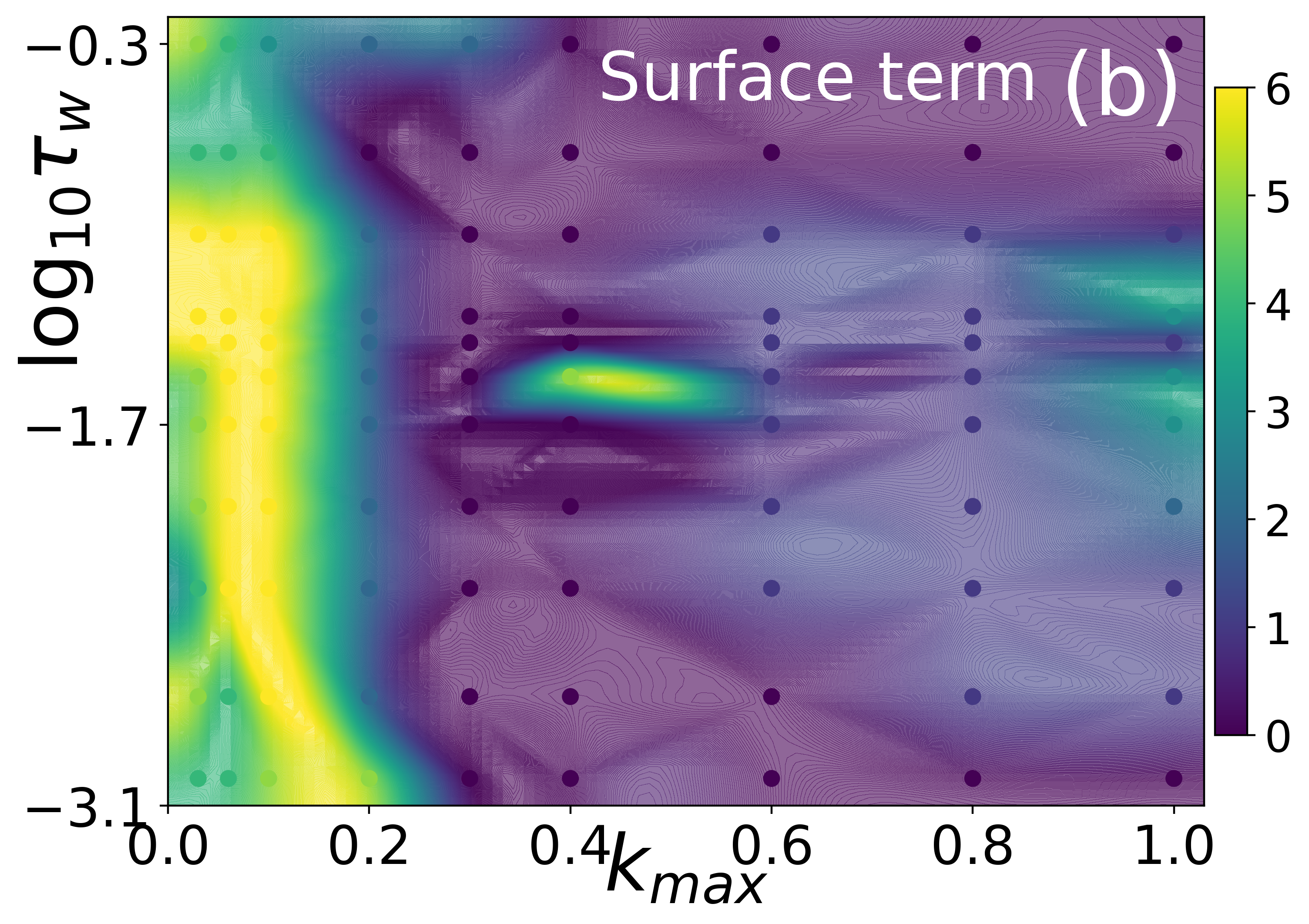}
     \includegraphics[width=0.49\linewidth]{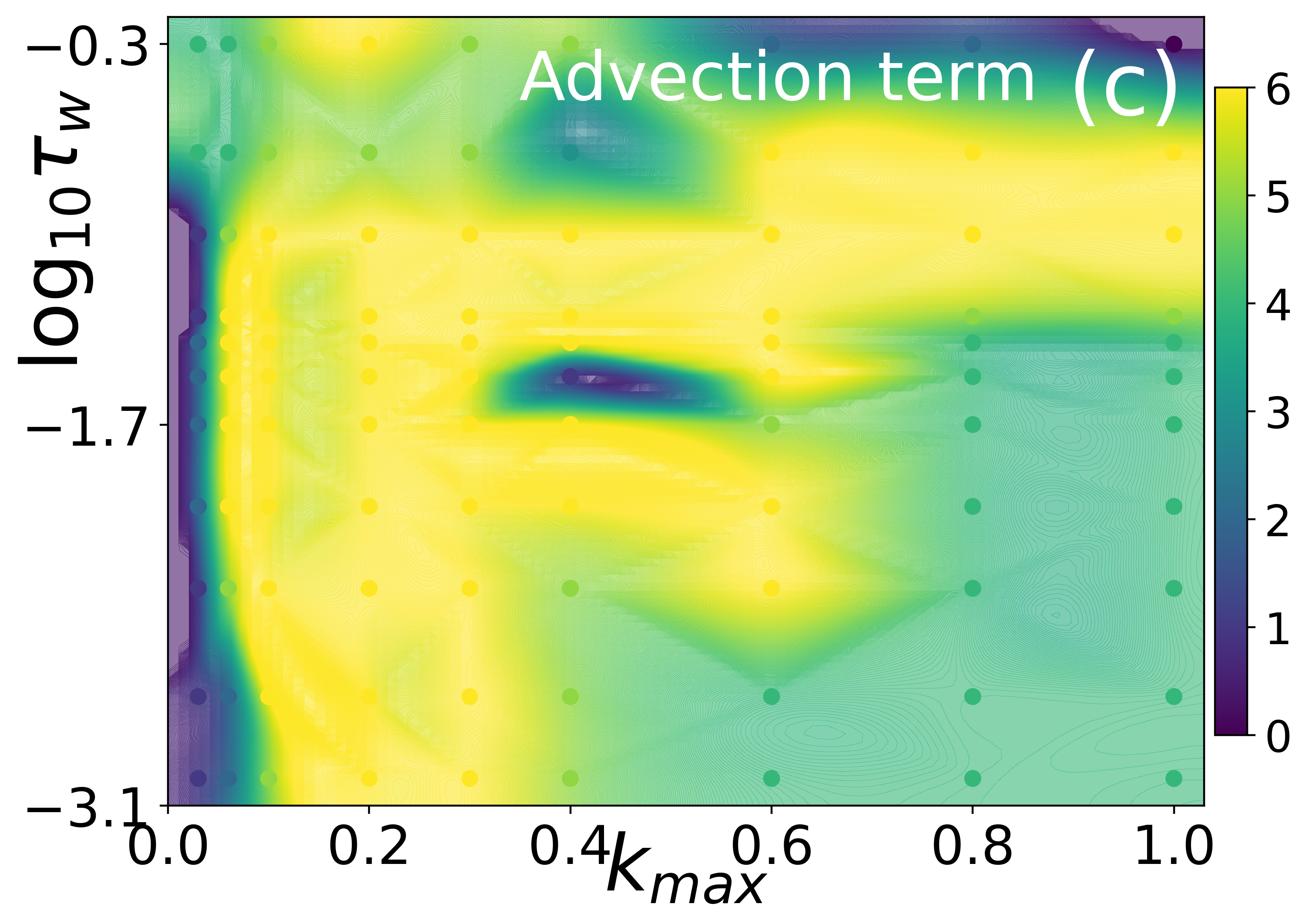}
     \includegraphics[width=0.49\linewidth]{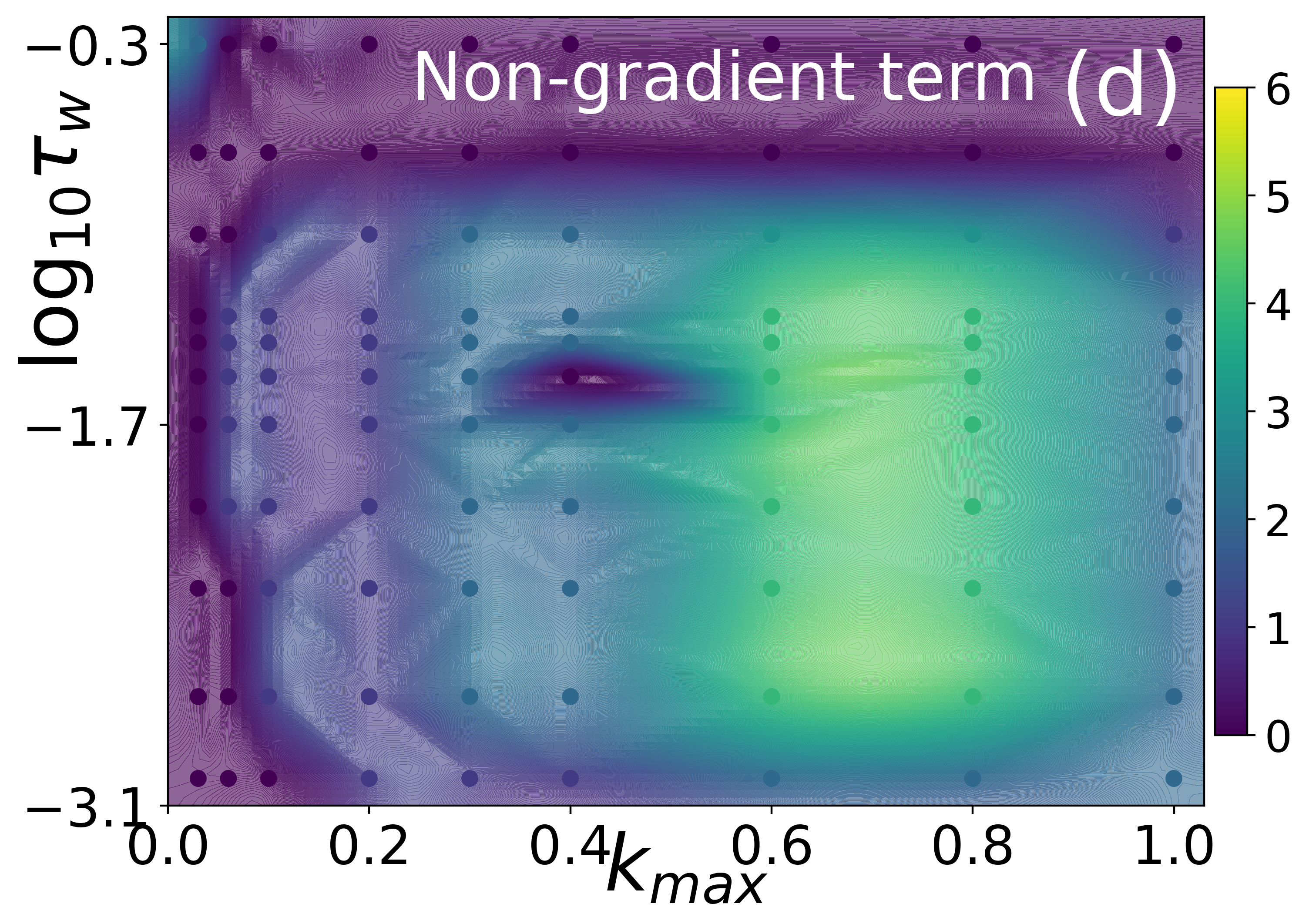}
  \caption{\label{fig.nonzerodictionary.ABP}(Color online)  Plot of the estimated total number of bulk terms (a), the total number of surface terms (b), the total number of advection terms (c), and the total number of Non-gradient terms (d) in each point of the parameter ($k_{\mathrm{max}},\tau_\mathrm{w}$) space.
  }
\end{figure}

We classify the dictionary terms in the hydrodynamic equation of density field into bulk terms ($\Delta \rho$, $\Delta \rho^2$, $\Delta \rho^3$,$\Pe \Delta \rho$, $\Pe \Delta \rho^2$, $\Pe \Delta \rho^3$), surface terms($\Delta^2 \rho$, $\Delta^2 \rho^2$, $\Delta^2 \rho^3$, $\Pe \Delta^2 \rho$, $\Pe \Delta^2 \rho^2$, $\Pe \Delta^2 \rho^3$), advection terms($\nabla \cdot \mathbf{p}$, $\nabla \cdot( \rho \mathbf{p})$,$\nabla \cdot ( \rho^2 \mathbf{p})$,$\Pe \nabla \cdot \mathbf{p}$, $\Pe \nabla \cdot ( \rho \mathbf{p})$,$\Pe \nabla \cdot ( \rho^2 \mathbf{p})$), and non-gradient terms ($\Delta |\mathbf{p}|^2$, \NY{$\nabla (|\mathbf{p}|^2 \mathbf{p})$}, $\nabla \cdot (\rho \nabla|\mathbf{p}|^2)$, $\nabla \cdot (|\mathbf{p}|^2 \nabla\rho)$, $\Pe \Delta |\mathbf{p}|^2$, \NY{$\Pe \nabla (|\mathbf{p}|^2 \mathbf{p})$}, $\Pe \nabla \cdot (\rho \nabla|\mathbf{p}|^2)$, $\Pe \nabla \cdot (|\mathbf{p}|^2 \nabla\rho)$).
Figure~\ref{fig.nonzerodictionary.ABP} shows the number of each term at different lengths and time scales.
We can see that the \NY{advection} terms appear in almost all scales, whereas \NY{the non-gradient terms preferentially appear in smaller length scales.
The bulk terms also appear in all scales, but their number is larger at the larger length scale.
The surface terms also appear at larger length scales.
Note that even at the larger length scales, the number of non-gradient terms is not zero. For example, in the case of $k_{\mathrm{max}}=0.1$ and $\tau_\mathrm{w}=0.02\tau$, which we discussed at the beginning of this section, one non-gradient term, which is $\Delta |\mathbf{p}|^2$, appears.
We speculate the following reasons for the appearance of each type of term, depending on space and time scales.
As we discussed above (and will discuss more quantitatively in Sec.\ref{sec.ABP.phase.diagram}), the phase separation occurs due to the balance of the advection and bulk terms.
At the smaller length scales, the spatial fluctuation of the density results in a larger amplitude of polarity density because $\mathbf{p} \sim \nabla \rho$ (in fact, $\mathbf{p} \sim \nabla \rho^2$ in the estimation results).
Therefore, the dynamics of $\rho$ is dominated by $|\mathbf{p}|^2$, which is included in the non-gradient terms. 
On the other hand, at the smaller $k_{\mathrm{max}}$, the spatial inhomogeneity becomes weaker, resulting from filtering the large $\mathbf{k}$ components.
}

\subsection{\NY{\BR{Evaluation} of the estimated results}}
\label{sec.ABP.phase.diagram}

\begin{figure}[ht]
    \includegraphics[width=0.99\linewidth]{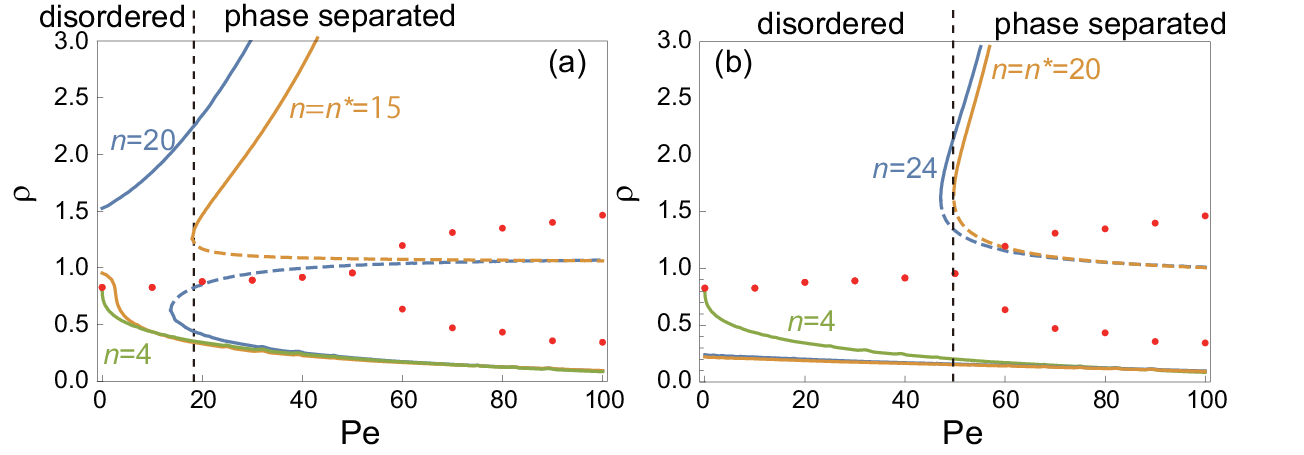}
  \caption{\label{fig.ABP.phase}(Color online) 
  Plot of local density vs $\Pe$ for the system density \NY{$\rho_0=0.826$ evaluated from the effective free energy Eq.\eqref{ABP.eff.freeenergy} using the estimated coefficients}. 
  \NY{
  The dictionary terms include $\alpha=\Pe$(a) and $\alpha=\Pe,\Pe^2$(b).
  }
  At smaller values of $\Pe$, the system is single-phase, and for increasing $\Pe$, it phase-separates. Dotted points \NY{(red)} are measured from the simulation of particle dynamics data.
\NY{The density at $\partial_{\rho} \tilde{f}_1=0$ using the estimated effective free energy for different \NY{number of dictionary terms are shown ($n=4$ (green), $15$ (orange), and $20$ (blue) in (a), and $n=4$ (green), $20$ (orange), and $24$ (blue) in (b)).
The optimal number of terms is $n^*=15$ for (a) and $n^*=20$ for (b).
}
Stable (unstable) states are shown in solid (dashed) lines.
The vertical line represents $\Pe_c$ estimated from the effective free energy.
}
}
\end{figure}

To analyze the estimation results further, we evaluate the effective free energy $\mathcal{F}[\rho]$ from the estimated coefficients.
We plug Eq.~\eqref{eq.ABP.polarity} into Eq.~\eqref{eq.ABP.density} to obtain the \NY{PDE} only for $\rho$.
Then, we rewrite the hydrodynamic equation as\NY{\cite{Tjhung:2018}}
\begin{align}
    \partial_t \rho&=
    \Delta \frac{\delta \mathcal{F} }{\delta \rho}
    + \mathcal{N}[\rho]
\end{align}
where $\mathcal{N}[\rho]$ is non-gradient terms, such as $\Delta |\nabla \rho|^2$.
The effective free energy is expressed as
\begin{align}
\mathcal{F} [\rho]
    &=
    \int d \mathbf{r}^2
    \left[
    \tilde{f}_1 (\rho(\mathbf{r}), \Pe)
    +
    \tilde{f}_2 (\rho(\mathbf{r}), \Pe) 
    |\nabla \rho(\mathbf{r})|^2
    \right].
    \label{ABP.eff.freeenergy}
\end{align}
Here, $\tilde{f}_1, \tilde{f}_2$ are \NY{polynomials in $\rho$ and $\Pe$}.
They can be computed from $f_1,f_2,f_3,g_1,g_2$ in Eq.~\eqref{eq.ABP.polarity} into Eq.~\eqref{eq.ABP.density}.
We focus on the bulk part of the effective free energy $\tilde{f}_1$, from which we can evaluate the coexistence line of the phase separation.
In this analysis, we estimate the hydrodynamic equation for the density field by replacing $\rho$ by $\rho - \rho_0$ in the dictionary terms.
We focus on the scales $\tau_\mathrm{w}=0.02\tau$ and $k_{\mathrm{max}}=0.1$ when the error is lowest.
\NY{
We performed the estimation using the dictionary terms up to linear order in $\Pe$, namely $\alpha=\Pe$ in Eq.~\eqref{eq.densityb} and Eq.\eqref{eq.polarity2b}.
This is the same as in Sec.~\ref{sec.result.ABP}.
In addition to the first case, we performed the second case in which we include the dictionary terms up to second order in $\Pe$, namely $\alpha=\Pe,\Pe^2$.
In the first and second cases, we obtained $n^*=15$ and $n^*=20$, respectively.
}
The estimated effective free energy reproduces two minima when \NY{the number of dictionary terms is large enough}. 
The minima of $\tilde{f}_1$ with respect to $\rho$ \BR{are} shown in Fig.~\ref{fig.ABP.phase} as a function of $\Pe$.
We also measure the histogram of local density in particle simulations of ABPs.
The points in Fig.~\ref{fig.ABP.phase} are the peak densities of the histogram.
\NY{
Although the coexistence of dense and dilute phases can be reproduced, the transition value of $\Pe$ deviates from the simulation result ($\Pe \approx 50$) when $\alpha=\Pe$.
The estimated transition value is $\Pe \approx 20$.
On the other hand, when $\alpha=\Pe, \Pe^2$, the transition value of $\Pe$ is close to the simulation result, $\Pe \approx 50$.
}
When we use fewer dictionary terms than the optimal $n$, the estimated effective free energy does not reproduce the coexistence of dense and dilute phases.
On the other hand, including more terms does not improve the result.
This analysis suggests that $n^*$ is determined so that the estimated hydrodynamic equation can reproduce the $\Pe$ dependence of the phase separation.
\NY{
Nevertheless, the actual values of dense and dilute density can not be quantitatively reproduced by the estimation results.
This is because the dictionary includes the terms only lower-order in $\rho$, which limits the expressibility of the effective free energy.
Including the higher-order terms in the dictionary does not solve this issue because those higher-order terms are strongly correlated with the lower-order terms and prevent reliable estimation.
}
We should stress that our method does \NY{not} estimate the free energy directly nor assume its existence.
We estimate the dynamical equations of the continuum fields from the data of particle dynamics.

\subsection{\NY{Time and length scales}}
\label{sec.abp.scales}

The ABP model has a time scale of translational diffusion $\tau = R^2/D$, which we chose as a unit time scale.
The time scale of rotational diffusion $D_r^{-1}=\tau/3$ is comparable to $\tau$.
The collision time scale is $\frac{R}{v_0} \sqrt{\frac{\pi}{\rho}} \approx \tau /\Pe$.
When the motility-induced phase separation occurs, first, many clusters appear at the time scale $\tau_{cl}$ and then each cluster grows and shows coarsening with the time scale $\tau_{gr}$ \cite{Redner:2013}.
During the coarsening process, the fraction of particles in the gas phase does not change.
The time scale $\tau_{cl}$ is dominated by the collision time scale.
With our choice of the system size,  $\tau_{gr} \approx 40 \tau$.
Our results of estimation suggest that its error is small when $\tau_\mathrm{w} \geq 10^{-2} \tau$, which is comparable to $\tau_{cl}$ for large $\Pe$.
This is consistent with the fact that the hydrodynamic equation describes phase separation.

\subsection{\NY{Generalizability}}

\NY{
We have checked the generalizability of the estimated model for the ABP model, similar to Sec.~\ref{sec.generalizability.VM}.
We have computed the errors for the density field using unseen data and the fixed estimated coefficients obtained by the training data. 
We obtain 
}$\NY{\tilde{E}_{\mathrm{test}}(n^*)}= 0.2868 \pm 0.0026$ with $\tau_w=0.02\tau$ and $k_{\rm max}=0.1$.
\NY{
This error is comparable to the training error, suggesting that our estimated hydrodynamic equation does not overfit the data.
}

\subsection{\NY{Sparsity}}
\label{sec.ABP.sparsity}

\NY{
Similar to Sec.~\ref{sec.discuss.sparsity}, we study how the estimation error in the ABP model decreases as we use more dictionary terms.
}
We found the three steps, $n^*_1$, $n^*_2$, and $n^*_3$ (Fig.~\ref{fig.error.num.terms.discussion}(b)).
After normalization, the error decreases in the same way for different $\tau_w$.
The first drop of the error includes $\Pe \nabla \cdot \mathbf{p}$, $\Delta\rho$, $\Delta \rho^2$, $Pe \Delta \rho^2$ and $Pe\nabla \cdot (\rho \mathbf{p})$ terms.
Those terms are independent of the choice of $\tau_w$.
Between $n_1^*$ and $n_2^*$, 7 terms discussed in  Sec.~\ref{sec.result.ABP} appear.
Again, these terms are independent of the choice of $\tau_w$.
On the other hand, the estimated terms between $n_2^*$ and $n_3^*$ are different for different $\tau_w$.
\NY{
These results are reasonable because, as we have seen, the error is dominated by the balance between the advection and bulk terms, which determines the instability of the uniform state.
Then, the instability is suppressed at a larger length scale by the surface terms, including higher-order spatial derivatives appearing between $n_1^*$ and $n_2^*$.
Also, the instability is tuned by the additional advection and bulk terms.
Between $n_2^*$ and $n_3^*$, the error is further decreased by fitting small-scale \BR{structures using} only the non-gradient terms.
However, the small-scale structures significantly include noise, and different terms can fit the remaining error.
As a result, the nonzero dictionary terms between $n_2^*$ and $n_3^*$ are different in the different conditions.
}

\begin{figure}[htb]
     \includegraphics[width=0.7\linewidth]{e-e0-win-k0.1_n.eps}
     \caption{\label{fig.error.num.terms.discussion.abp}(Color online) 
     \NY{
The plot of the error, $\tilde{E}(n)-\tilde{E}_{\mathrm{min}}/(1-\tilde{E}_{\mathrm{min}})$ vs the number of estimated terms $n$ for \NY{ABPs} under different time windows $\tau_\mathrm{w}$ with $k_{\rm max}=0.1$.
The error is normalized so that its minimum value is zero for all $\tau_\mathrm{w}$.
  }
  }
\end{figure}

\section{Discussions and Conclusion}
\label{sec.discussion}

To summarize, we propose a method to estimate hydrodynamic equations from data of microscopic particle dynamics.
We transform the data of the trajectory of individual particles into field data with the Gaussian kernel. 
Then, we apply spatial and temporal filters to the field data so that we obtain the coarse-grained data at different lengths and time scales.
Using the method of estimation of \NY{PDEs} with encouraging sparsity, we can estimate the hydrodynamic equations corresponding to the coarse-grained data.
As we stressed in the Introduction, the purpose of estimating the hydrodynamic equations is not to reproduce all the microscopic dynamics of particle simulations.
It is instead to extract qualitative features hidden in complex microscopic dynamics.
Our results suggest that the estimated hydrodynamic equations depend on the length and time scales we are looking at.

\NY{
We demonstrate that our estimation method works both for the Vicsek and ABP models, namely, their phase diagrams can be reproduced, at least semi-quantitatively, by the estimated hydrodynamic equations.
Still, we have to extend our method to improve quantitative predictive power. 
Among many possibilities, we believe a choice of dictionary terms is important.
Although we construct the dictionary terms based on the polynomial expansions, this is possibly not the best way.
Because the higher-order terms strongly correlate with the lower-order terms, including the higher-order terms does not necessarily improve the estimation.
On the other hand, to reproduce accurate physical properties, such as an effective free energy, the lower-order terms in the polynomial expansion are not enough.
Instead, we may consider other \BR{bases}, suitable for the problem at hand.
We will explore this direction in the future.
}

\NY{
Our method should work for the system whose hydrodynamic description is expressed by PDEs.
There are many potential applications of the variants of the Vicsek model and ABPs to more realistic soft materials and biological systems.
Even for the ABPs, it has been studied that hexatic and solid phases appear at high volume fraction\cite{Digregorio:2018}.
For the Vicsek model, small modifications of the model would result in different symmetries\cite{Chate:2020}.
Estimating those systems requires additional \BR{variables}, such as local hexatic order, but the hydrodynamic equations of those variables are an interesting future direction.
}

\begin{acknowledgments}
The authors acknowledge the support from the JSPS KAKENHI Grant
number JP20K03874, 24H02203, and JP24K00591 to N.Y. 
This work is also supported by the JST FOREST Program Grant Number JPMJFR2140 to N.Y.
\NY{
N.Y. would like to thank the Isaac Newton Institute for
Mathematical Sciences, Cambridge, for support and hospitality during the SPL programme, funded by EPSRC
grant no EP/R014604/1.
}
\end{acknowledgments}

\appendix

\section{Appendixes}

\subsection{Dictionary terms}
\NY{
We use the following dictionary terms as a standard method if otherwise stated.
For the Vicsek model, $\alpha=\eta$ and for the ABP, $\alpha=\Pe$.
When we include additional dictionary terms, such as $\alpha=\Pe^2$, or some dictionary terms are replaced by others, for example, $\alpha=\exp(-\eta^2/2)$ instead of $\alpha=\eta$, we state it explicitly in the text.
}

\NY{
For the density field in Eq.~\eqref{eq.density}, the following $26$ dictionary terms are used with corresponding coefficients from $a_1$ to $a_{26}$ (see Eq.\eqref{eq.Vicsek.density} for the Vicsek model and Eq.\eqref{eq.ABP.density} for the ABP):
}
\begin{align}
    C_l =&
    \Big(
    \nabla \cdot \mathbf{p}, \Delta \rho, \nabla \cdot (\rho \mathbf{p}), \Delta \rho^2, \Delta |\mathbf{p}|^2, \nabla \cdot (\rho^2 \mathbf{p}), \Delta \rho^3,
    \nonumber \\ &\nabla \cdot (|\mathbf{p}|^2 \mathbf{p}), \nabla \cdot (\rho \nabla |\mathbf{p}|^2), \nabla \cdot (|\mathbf{p}|^2 \nabla \rho), \Delta^2 \rho, \Delta^2 \rho^2,
    \nonumber \\ &  \Delta^2 \rho^3, \alpha\nabla \cdot \mathbf{p}, \alpha\Delta \rho, \alpha\nabla \cdot (\rho \mathbf{p}), \alpha\Delta \rho^2, \alpha\Delta |\mathbf{p}|^2, 
    \nonumber \\ &  \alpha\nabla \cdot (\rho^2 \mathbf{p}), \alpha\Delta \rho^3, \alpha\nabla \cdot (|\mathbf{p}|^2 \mathbf{p}), \alpha\nabla \cdot (\rho \nabla |\mathbf{p}|^2),  
    \nonumber \\ & \alpha\nabla \cdot (|\mathbf{p}|^2 \nabla \rho), \alpha\Delta^2 \rho, \alpha \Delta^2 \rho^2,
        \alpha\Delta^2 \rho^3
        \Big)    \label{eq.densityb}
\end{align}

\NY{
For the polarity density field in Eq.~\eqref{eq.polarity}, the following \BR{$20$} dictionary terms are used with corresponding coefficients from $b_1$ to $b_{20}$ (see Eq.\eqref{eq.Vicsek.polarity} for the Vicsek model):
}
\begin{align}
   D_l
    =& 
    \Big(\mathbf{p}, \rho \mathbf{p}, \rho^2 \mathbf{p}, |\mathbf{p}|^2 \mathbf{p}, \nabla \rho, (\mathbf{p} \cdot \nabla)\mathbf{p}, \nabla(\nabla \cdot \mathbf{p}), \Delta \mathbf{p},     
    \nonumber \\ &  \nabla |\mathbf{p}|^2, (\nabla \cdot \mathbf{p})\mathbf{p}, \alpha\mathbf{p}, \alpha\rho \mathbf{p}, \alpha\rho^2 \mathbf{p}, \alpha|\mathbf{p}|^2 \mathbf{p}, \alpha\nabla \rho, 
     \nonumber \\ & \alpha(\mathbf{p} \cdot \nabla)\mathbf{p}, \alpha\nabla(\nabla \cdot \mathbf{p}), \alpha\Delta \mathbf{p},  
     \alpha\nabla |\mathbf{p}|^2, \alpha(\nabla \cdot \mathbf{p})\mathbf{p}
     \Big),
     \label{eq.polarityb}
\end{align}

\NY{
For the polarity density field in Eq.~\eqref{eq.polarity2}, the following $22$ dictionary terms are used with corresponding coefficients from $\tilde{b}_1$ to $\tilde{b}_{22}$ (see Eq.\eqref{eq.ABP.polarity} for the ABP and description above Eq.~\eqref{eq.polarity2}):
}
\begin{align}
     \tilde{D}_l
     =&
         \Big(\nabla\rho, \nabla\rho^2, \nabla\rho^3, 
    \nabla (\Delta \rho), \nabla(\Delta \rho^2), \nabla (\Delta \rho^3), 
     \nonumber \\
    &  
    \nabla |\nabla\rho|^2, (\nabla \rho)(\Delta\rho), \nabla \rho|\nabla \rho|^2, \rho (\nabla \rho) (\Delta\rho), \rho \nabla |\nabla\rho|^2, 
    \nonumber \\
    &
     \Pe\nabla\rho, \Pe\nabla\rho^2, \Pe\nabla\rho^3, \Pe\nabla (\Delta \rho),
     \Pe\nabla(\Delta \rho^2), 
     \nonumber \\
    &
    \Pe\nabla (\Delta \rho^3), \Pe\nabla |\nabla\rho|^2, \Pe(\nabla \rho)(\Delta\rho), \Pe\nabla \rho|\nabla \rho|^2, 
    \nonumber \\
    &
    \Pe \rho (\nabla \rho) (\Delta\rho), \Pe \rho \nabla |\nabla\rho|^2
     \Big),
    \label{eq.polarity2b}
\end{align}

\subsection{Coefficients in Eqs.\eqref{eq.ABP.density} and \eqref{eq.ABP.polarity}}
\label{appendix.ABP.coeff}

\begin{align}
f_1
=&
-0.887\Pe 
-0.492 \rho \Pe
+ 0.370 \rho^2 \Pe
\nonumber \\
&
+ 14.538
-27.455 \rho
+ 13.692 \rho^2
\\
f_2=& 794.008 -998.452 +430.677
\nonumber \\& 
-3.359\Pe + 2.235\Pe + 0.226 \Pe
\\
f_3
=& 22969.929 -20796.103 + 5533.872
\nonumber \\&  -184.109 \Pe + 152.673 \Pe - 34.304 \Pe
\\
g_1
=& -10.422 + 7.784 + 0.0049 \Pe -0.0029 \Pe
\\
g_2
=& -454.836 + 275.210
\end{align}

\nocite{*}

%

\end{document}